 \documentclass[10pt]{IEEEtran}
\usepackage[a4paper]{geometry}
\IEEEoverridecommandlockouts

\usepackage{cite}
\usepackage{amsmath,amssymb,amsfonts}
\usepackage{graphicx}
\usepackage{textcomp}
\usepackage{xcolor}
\usepackage{float}
\usepackage{array}
\usepackage{algorithm}
\usepackage{algpseudocode}
\usepackage{bm}
\usepackage{tabularx}
\usepackage{multirow}
\usepackage{booktabs}
\usepackage{stfloats}
\usepackage{ntheorem}

\usepackage{geometry}
\usepackage{subfigure}
\usepackage{stfloats}
\makeatletter
\renewcommand{\fnum@algorithm}{}

\usepackage{epsfig}
\usepackage{epstopdf}

\begin{document}
\title{
	Dual-Mapping Sparse Vector Coding for Phase Noise-Resilient Short-Packet Transmission}
\author{{Yanfeng~Zhang,~\IEEEmembership{\normalsize{Member,~IEEE}},}
Xi'an Fan,
        Xu~Zhu,~\IEEEmembership{\normalsize{Senior Member,~IEEE}},~Yujie Liu,
Jinkai Zheng,~Weiwei Yang,\\
Wen Wu,~\IEEEmembership{\normalsize{Senior Member,~IEEE}\vspace{-0pt}},~Tom H. Luan,~\IEEEmembership{\normalsize{Fellow,~IEEE}},~and Yong Liang Guan,~\IEEEmembership{\normalsize{Senior Member,~IEEE}\vspace{-0pt}}
\vspace{-10pt}
\thanks{Part of this work has been presented at IEEE GLOBECOM 2025 \cite{ZhangDMSVC2025}.}
\thanks{Yanfeng Zhang, Jinkai Zheng and Weiwei Yang are with the School of Electrical Engineering and Intelligentization, Dongguan University of Technology, Dongguan, China. (e-mails: yfzhang@ieee.org; jkzheng@ieee.org; yangww@dgut.edu.cn)}
\thanks{Xi'an Fan is with the School of Computer Science and Technology, Dongguan University of Technology, Dongguan, China. (e-mail: exianfan@gmail.com)}
\thanks{Xu Zhu is with the School of Information Science and Technology, Harbin Institute of Technology, Shenzhen 518055, China. (e-mail: xuzhu@ieee.org)}
\thanks{Yujie Liu and Yong Liang Guan are with Continental-NTU Corporate Lab, Nanyang Technological University, Singapore. (e-mail: yujie.liu@ieee.org, eylguan@ntu.edu.sg.)}
\thanks{Wen Wu is with the Department of Strategic and Advanced Interdisciplinary Research, Pengcheng Laboratory, Shenzhen 518000, China. (e-mail: wuw02@pcl.ac.cn)}
\thanks{Tom H. Luan is with the School of Cyber Science and Engineering, Xi’an Jiaotong University, Xi’an, China. (e-mail: tom.luan@xjtu.edu.cn)}
\vspace{-0pt}
}

\maketitle

\begin{abstract}
Sparse vector transmission (SVT) has emerged as a promising technique for ultra-reliable low-latency short-packet communications. However, existing SVT schemes typically assume negligible phase noise (PN), an assumption that rarely holds in practical wireless systems. In this paper, a dual-mapping sparse vector coding (DM-SVC) scheme is proposed for short-packet communications subject to PN. In DM-SVC, pilot symbols are mapped onto multiple non-zero blocks and data symbols onto isolated non-zero elements within a single sparse vector, thereby enabling pilot–data separation through distinct sparsity patterns rather than explicit resource partitioning. Moreover, the indices of pilot blocks convey additional information bits, further improving spectral efficiency. A basis expansion model is adopted to represent the PN process, substantially reducing the number of parameters to be estimated. Furthermore, an iterative joint PN estimation and data decoding algorithm is developed, where pilot block indices are first detected exploiting block-sparse priors, after which PN estimation and data decoding proceed iteratively. Simulation results show that DM-SVC could achieve block error rate performance close to that of perfect PN compensation, while offering improved spectral efficiency and reduced codebook storage overhead compared to state-of-the-art SVT schemes.
\end{abstract}
\vspace{-5pt}

\vspace{-0pt}
\section{Introduction}
Ultra-reliable and low-latency communication (URLLC) is a key enabler in 5G and emerging 6G wireless systems, targeting latency-critical applications such as autonomous driving \cite{DMoya2022}, industrial Internet of Things (IIoT) \cite{DNguyen2022}, and remote sensing control \cite{Liufan2022JSAC}. These services typically demand a reliability of 99.999\% or above while constraining the end-to-end latency to 1 ms or less \cite{CYueWCM2023}. A straightforward approach to reducing latency is to shorten the packet to a few tens of bytes or less \cite{SheChangyang2018}. However, short packets severely limit the achievable coding gain, making reliable transmission considerably more challenging. Conversely, improving reliability typically requires additional redundancy, whether through retransmissions or extra parity bits, both of which inevitably increase transmission latency. Conventional long-packet transmission schemes therefore struggle to meet the stringent requirements of URLLC, motivating the development of new transmission techniques tailored to short-packet communications.

Recently, a new class of transmission scheme called sparse vector transmission (SVT) \cite{Kim20201} has attracted increasing research interest. In SVT, information bits are encoded in the index positions of a high-dimensional sparse vector, which is then compressed through a non-orthogonal codebook and transmitted over time–frequency resources. Unlike conventional schemes that rely on error-correction codes, SVT receiver achieves reliable decoding by identifying the indices of a small number of non-zero elements from the received signal. Extensive studies have demonstrated that SVT outperforms traditional coding techniques in the finite-blocklength regime in terms of both reliability and latency. Consequently, SVT is regarded as a promising transmission technique for URLLC.

\vspace{-10pt}
\subsection{Related Work}

Conventional error-correcting codes such as low-density parity-check (LDPC) codes and polar codes perform well at long block lengths. However, their decoding performance degrades markedly in the short-packet regime. Specifically, LDPC codes suffer from insufficient coding gain due to limited iterative decoding \cite{MShirvanimoghaddamWCM2019}, while polar codes are hampered by incomplete channel polarization at short block \mbox{lengths \cite{QZhang2017WCL}.} In addition, forward error correction (FEC) codes have also been investigated for short-packet \mbox{communications \cite{Dizhang2026}.} Meanwhile, a careful balance among reliability, decoding complexity, and latency is required for FEC schemes in the short-packet regime. SVT has emerged as a promising alternative for short-packet transmission in URLLC systems. Existing work on SVT can be broadly categorized into three aspects: encoding design, decoding algorithms, and extensions to various communication scenarios.

The sparse regression codes (SPARCs) \cite{AJosephTIT2012} and sparse vector coding (SVC) \cite{Ji2018} are the two primary approaches to SVT encoding design. The original SPARCs scheme partitions a sparse vector into multiple sections and places one non-zero element per section, with information bits encoded through both section and non-zero element indices. The SVC scheme extends this concept by mapping information bits to the positions of non-zero elements without section partitioning, offering greater design flexibility. Existing studies have primarily focused on improving spectral efficiency (SE) and optimizing encoding structure. To improve SE, modulated symbols are assigned to non-zero elements, leading to a family of modulated SVT schemes, such as enhanced \mbox{SVC (ESVC) \cite{Kim20202},} sparse superposition codes (SSC) \cite{ZhangXuewan2022}, block orthogonal sparse superposition (BOSS) codes \cite{DhanTWC2023}, and generalized SPARCs (GSPARCs) \cite{MSinhaTCOM2024}. In addition, new information-bearing dimensions have been explored, including the index resources of multiple constellations with different modulation modes \cite{YangLinjie2024} and the interleaving pattern in hyper-dimensional modulation \cite{CHsuTCOM2023}. The block SVC (BSVC) scheme proposed in \cite{yfzhang2025BSVC} maps information bits to nonzero block indices rather than individual nonzero-element indices, thereby inducing a block-sparse structure that reduces the codeword length and improves SE. To reduce the length of sparse vectors, several encoding strategies have been proposed, including index redefinition \cite{Zhangxue25}, sparse vector extension \cite{SparseExtendedSVC25}, and one-hot vector fusion \cite{OnehotSVC25}. The block error rate (BLER) performance of SVC can be further improved by optimizing the codebook matrix to reduce column correlation \cite{Yang2023}. Furthermore, conventional polar codes \cite{KuvshinovAleksey2024} and LDPC \mbox{codes \cite{EbertJamison2025}} have been concatenated with SPARCs, achieving low-complexity decoding with competitive BLER performance.

The SVT decoding process involves identifying the indices of non-zero elements from noisy channel observations. Various decoding algorithms have been proposed for SVT systems. The maximum likelihood decoder ensures arbitrarily low error probability at rates below channel capacity as code length increases, providing a theoretical performance benchmark for SPARCs \cite{AJosephTIT2012}. To reduce complexity, the approximate message passing (AMP) algorithm has been applied to SPARCs \cite{CRushTIT2021}, enabling low-complexity iterative decoding under large-system assumptions. For SVC, greedy algorithms such as multipath matching pursuit (MMP) \cite{Ji2018} and matching pursuit \cite{MSinhaTCOM2024} are more commonly adopted owing to their favorable complexity–performance trade-off. To exploit structured sparsity, a cyclic block orthogonal matching pursuit algorithm has been developed for BSVC \cite{yfzhang2024BSVC}. A parallel match-and-decode algorithm has been proposed in \cite{MSinhaTCOM2024} to improve the decoding performance of GSPARCs. Recently, data-driven decoding methods have been explored for SVT schemes. \mbox{In \cite{CBian2023TCOM},} a neural network-based encoder has been developed to learn the mapping between received signals and sparse supports, aiming to improve decoding performance under complex channel conditions.

The SVT scheme has been extended to a wide range of communication scenarios. In multi-user systems, SVT has been studied in multiple-input multiple-output (MIMO) systems \cite{ZhangRuoyu2021}, non-orthogonal multiple access (NOMA) \mbox{systems \cite{Sabapathy2023}}, grant-free access systems \cite{Luoyingzhe24}, and rate-splitting multiple access (RSMA) systems \cite{MultiuserSVC25}. These works demonstrate the potential of SVT to support low-latency transmission. 
For communication systems with practical hardware constraints, SVT schemes have been developed for systems with low-resolution analog-to-digital \mbox{converters \cite{ZYF2024ICCC}} and memory-limited devices \cite{ZYFWCNC25}. Additionally, SVC has been applied to challenging communication environments such as high-mobility scenarios \cite{ZhangYf2023}. Recent work has explored SVT in emerging paradigms including semantic communications \cite{Zhanxunyang25}, demonstrating the broad applicability of SVT across diverse communication systems.

\vspace{-0pt}
\subsection{Motivations and Contributions}

As discussed above, most existing SVT schemes focus on encoding and decoding design or extensions to various communication scenarios. These works commonly assume ideal phase synchronization or negligible phase noise (PN). Under this assumption, sparse recovery based decoders can reliably identify the support of sparse vectors. In practical systems, however, PN is unavoidable and introduces time-varying phase distortions across transmitted symbols. This effect breaks the assumed sparse signal model and causes a sharp degradation in support detection accuracy. The problem is more severe in short-packet transmission, where limited redundancy cannot compensate for model mismatch. A straightforward solution is to introduce dedicated pilot symbols for PN estimation and compensation, but this approach increases signaling overhead and transmission latency. This motivates the development of an SVT framework that explicitly accounts for PN while enabling joint PN estimation and data decoding.

To address the above problem, a dual-mapping SVC (DM-SVC) scheme is proposed, where pilots and data are mapped onto a sparse vector through two distinct sparse patterns. The main contributions are summarized as follows\footnote{It is noteworthy that this paper is a substantive extension of our previous work \mbox{in \cite{ZhangDMSVC2025}.} The main differences lie in that: 1) the work in \cite{ZhangDMSVC2025} focuses on short-packet transmission without considering PN, whereas this paper extends the dual-mapping SVT (DM-SVT) framework to PN-impaired scenarios; 2) BEM-based PN modeling and estimation are studied in this paper to reduce the pilot overhead, while PN modeling and estimation are not considered in \cite{ZhangDMSVC2025}; 3) the proposed DM-SVC scheme employs the values of non-zero block elements as pilot symbols for PN estimation, whereas both the non-zero block indices and non-zero element indices in \cite{ZhangDMSVC2025} are used only for data transmission.}.

\begin{itemize}
\item The DM-SVC scheme is proposed for PN-impaired short-packet communication systems. Specifically, the pilot symbols are mapped onto multiple non-zero blocks, while data symbols are mapped onto isolated non-zero elements within a single sparse vector. This design enables pilot–data separation through distinct sparsity patterns, without resorting to time, frequency, or space division multiplexing. The pilot blocks and data symbols are superimposed over the same transmission resources, thereby avoiding the allocation of dedicated pilots for PN estimation, as required in conventional schemes. Furthermore, a basis expansion model (BEM) is adopted to represent the PN process through a small number of coefficients, substantially reducing the number of unknown parameters and enabling efficient PN estimation in the short-packet regime.

\item A dynamic pilot deployment strategy is developed for DM-SVC, in which the positions of non-zero pilot blocks vary across transmissions and are unknown to the receiver. The pilot symbols within the non-zero blocks serve as known references for PN estimation, while the block indices simultaneously conveys additional information bits. Unlike conventional pilot designs, in which pilot symbols occupy resources without contributing to data transmission, DM-SVC reuses pilot block indices as an additional information-bearing dimension. This gain requires neither higher-order modulation, longer code lengths, nor increased sparsity, but instead exploits the structural difference between block-sparse pilot components and element-sparse data components. This addresses a key limitation of existing SVC schemes, which often sacrifice SE for high reliability.

\item An iterative algorithm for joint PN estimation and data decoding is developed to exploit the dual-sparse structure of DM-SVC. In each iteration, a cyclic block matching pursuit (CBMP) algorithm is designed to identify the active pilot blocks by leveraging block-sparsity priors, namely the number of blocks and the block length. The PN is then estimated by recovering a small number of BEM coefficients from the detected pilot blocks. With the updated PN estimate, the algorithm detects the support of the non-zero data elements and subsequently demodulates the corresponding symbols. Moreover, data symbols decoded in earlier iterations are further used as pseudo-pilots to refine the PN estimate in subsequent iterations. This iterative mechanism improves both PN estimation and data decoding. Extensive simulation results demonstrate that the proposed decoding algorithm outperforms existing schemes in terms of BLER, with fast convergence.
\end{itemize}

The rest of this paper is organized as follows. In Section II, the DM-SVC encoding with dual-mapping pattern is presented. In Section III, the iterative joint PN estimation and data decoding algorithm is described. In Section IV, the complexity, SE and codebook storage overhead of DM-SVC are analyzed. Extensive simulation results are shown in Section V, while conclusions are drawn in Section VI.

\emph{Notations:} Bold symbols represent vectors or matrices. ${( \cdot )^{\rm T}}$, ${( \cdot )^{\rm H}}$ and ${( \cdot )^{ \dag }}$ denote the transpose, conjugate transpose and matrix inversion, respectively. $|| \cdot |{|_p}$ denotes ${\ell _p}$ norm operation. $\left\lfloor  \cdot  \right\rfloor$ is the round-down operation. ${\tbinom{N}{K}}$ denotes the number of combinations of selecting $K$ items from $N$ items. ${\rm{circ}}\{\bf x\}$ is a circular matrix generated with $\bf x$. $\odot$ denotes the Hadamard (element-wise) product.

\vspace{-10pt}
\section{Encoding of DM-SVC}
This section presents the DM-SVC encoding process, covering the two sparse mapping patterns, codebook construction, and the input–output signal model.

\vspace{-5pt}
\subsection{Preliminaries of Conventional SVC}
The key idea of the conventional SVC scheme is to embed information in both the indices and the non-zero values of a length-$N$ sparse vector $\mathbf s$. Consider a transmission block carrying $b$ bits. The encoder first partitions these bits into two groups: $b_{\rm I}$ bits for index selection and $b_{\rm S}$ bits for symbol modulation, satisfying $b = b_{\text{I}} + b_{\text{S}}$. The first group determines which $K$ out of $N$ positions are activated, yielding $\binom{N}{K}$ possible combinations and thus encoding $b_{\text{I}} = \lfloor \log_2 \binom{N}{K} \rfloor$ bits. The second group assigns a quadrature amplitude modulation (QAM) symbol from an $M_{\rm S}$-ary constellation $\mathcal{A}$ to each of the $K$ active positions, contributing $b_{\text{S}} = K \log_2 M_{\rm S}$ bits. For instance, with $N=16$, $K=4$, and $M_{\rm S}=4$, the sparse mapping conveys $b_{\rm I} = 10$ bits and the symbol modulation contributes $b_{\rm S} = 8$ bits, yielding 18 bits per sparse vector. After sparse mapping, the non-zero elements of $\mathbf{s}$ are spread over $M$ time–frequency resources through a codebook matrix $\mathbf{G}$. The spreading process can be expressed as
\begin{equation}
\mathbf{x} = \mathbf{G}\mathbf{s},
\end{equation}
where $\mathbf{x} \in \mathbb{C}^M$ is the transmitted signal.

To improve decoding performance, an angle rotation is typically applied to the modulated symbols. $\mathcal{A} = \{{\mathcal{A}}_1, {\mathcal{A}}_2, \cdots, {\mathcal{A}}_{M_{\text{mod}}}\}$ denote the $M_{\text{S}}$-ary QAM constellation. The $k$-th rotated symbol is then given by $s_k = e^{j\theta_k} {\mathcal{A}}$, where the rotation angle satisfies
\begin{equation}
\theta_k \in \left\{\theta_n = \frac{\pi}{2K}(n-1), n = 1, 2, \cdots, K\right\}.
\end{equation}

The spread signal $\mathbf{x}$ is then transformed to the time domain by an inverse discrete Fourier transform (IDFT), yielding $\mathbf{x}_{\text{T}} = \mathbf{F}^{\text{H}} \mathbf{x}$, where $\mathbf{F} \in \mathbb{C}^{M \times M}$ is a normalized DFT matrix. A cyclic prefix (CP) of length $L_{\text{CP}}$ is prepended to $\mathbf{x}_{\text{T}}$ to mitigate inter-symbol interference.At the receiver, after CP removal and DFT processing, the frequency-domain received signal is given by
\begin{equation}
\mathbf{y} = \mathbf{F} \mathbf{H}_{\text{T}} \mathbf{F}^{\text{H}} \mathbf{x} + \mathbf{w} = \mathbf{H}_{\text{F}} \mathbf{x} + \mathbf{w},
\end{equation}
where $\mathbf{H}_{\text{T}} \in \mathbb{C}^{M \times M}$ and $\mathbf{H}_{\text{F}} \in \mathbb{C}^{M \times M}$ are time-domain and frequency-domain channel matrices, respectively. $\mathbf{w} \sim \mathcal{CN}(0, \sigma^2 \mathbf{I}_M)$ denotes complex additive white Gaussian noise (AWGN) vector with variance $\sigma^2$. Under quasi-static fading, the channel remains constant over the transmission duration, so that $\mathbf{H}_{\text{T}}$ exhibits a circulant structure and $\mathbf{H}_{\text{F}}$ reduces to a diagonal matrix.

\vspace{-5pt}
\subsection{Proposed DM-SVC Scheme}
\emph{1) Sparse Mapping Pattern of DM-SVC}

Consider an OFDM-based short-packet transmission framework in which each packet carries $b$ information bits. As illustrated in Fig. 1, the input bit stream is partitioned into three groups, i.e., block index bits, single-element index bits, and symbol bits. Specifically, $b_{\mathrm{P,I}}$ bits are assigned to determine the position of $K_{\mathrm{P}}$ non-zero blocks via block sparse mapping, forming the block index set $\mathcal{B}$. Then, $b_{\mathrm{D,I}}$ bits are used to specify the locations of isolated $K_{\mathrm{D}}$ non-zero elements through single-element sparse mapping, resulting in the data index set $\mathcal{S}$. The remaining $b_{\mathrm{D,S}}$ bits are mapped onto $M_{\mathrm{S}}$-ary QAM symbols.

The number of bits conveyed by each group is determined by the corresponding sparse mapping pattern and the modulation order. The number of bits encoded by the pilot block index pattern is given by
\begin{equation}
b_{\mathrm{P,I}} = \left\lfloor \log_2 \binom{N -  K_{\mathrm{P}}(L_{\mathrm{P}} - 1)}{K_{\mathrm{P}}} \right\rfloor,
\end{equation}
where $K_{\mathrm{P}}$ denotes the number of non-zero pilot blocks and $L_{\mathrm{P}}$ is the block length. The number of bits mapped to the indices of isolated non-zero data elements is expressed as
\begin{equation}
b_{\mathrm{D,I}} = \left\lfloor \log_2 \binom{N - 2K_{\mathrm{P}} - K_{\mathrm{P}} L_{\mathrm{P}}}{K_{\mathrm{D}}} \right\rfloor,
\end{equation}
where $K_{\mathrm{D}}$ represents the number of isolated non-zero data elements. The number of bits conveyed by constellation modulation is
\begin{equation}
b_{\mathrm{D,S}} = K_{\mathrm{D}}\log_2( {M_{\mathrm{S}})}.
\end{equation}

To illustrate the proposed bit mapping strategy, consider a sparse vector of length $N = 21$ with $K_{\mathrm{P}} = 1$ pilot block of length $L_{\mathrm{P}} = 2$ and $K_{\mathrm{D}} = 2$ isolated non-zero data elements. The pilot block index pattern conveys $b_{\mathrm{P,I}} = \left\lfloor \log_2 \binom{N -  K_{\mathrm{P}}(L_{\mathrm{P}} - 1)}{K_{\mathrm{P}}} \right\rfloor = 4$ bits. After allocating the pilot block and guard elements, the data element index pattern encodes
$b_{\mathrm{D,I}} = \left\lfloor \log_2 \binom{N - 2K_{\mathrm{P}} - K_{\mathrm{P}} L_{\mathrm{P}}}{K_{\mathrm{D}}} \right\rfloor  = 7$ bits. The QAM symbols contribute $b_{\mathrm{D,S}} = K_{\mathrm{D}} M_{\mathrm{S}} = 4$ bits. 

\begin{figure*}[htbp]
	\centerline{\includegraphics[width=0.68\textwidth]{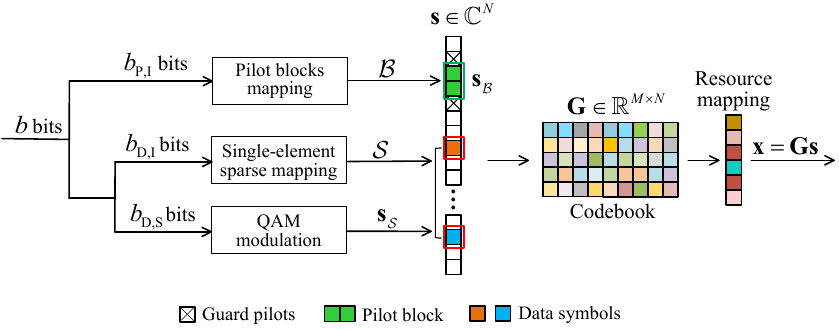}}
	\caption{Diagram of the encoding process of DM-SVC scheme. }
	\label{fig1}
\vspace{-0pt}
\end{figure*}

The dual mapping strategy is motivated by two considerations. First, the non-zero blocks serve as embedded pilots for PN estimation, and their block-sparse structure facilitates reliable recovery of the PN coefficients. Second, pilot and data components are distinguished solely by their sparsity patterns, allowing the receiver to decouple PN estimation from data decoding without dedicated pilot overhead.

\emph{Remark 1}:
In conventional dedicated-pilot transmission schemes, pilots and data are transmitted over different time or frequency resources, which results in high transmission latency and low SE. Such scheme is not well suited for low-latency communications. Conventional superimposed pilots-based schemes allow pilots and data to share the same resources, yet the pilot symbols serve only for PN estimation and carry no additional information. In contrast, the proposed DM-SVC scheme uses the pilot symbols within non-zero blocks for PN estimation while simultaneously exploiting the block indices to convey additional information bits, thereby improving SE without extra transmission overhead.

\emph{Remark 2}: In the proposed DM-SVC scheme, two guard pilots are inserted around each non-zero pilot block to ensure reliable separation between non-zero pilot blocks and isolated non-zero data elements. Without such guard pilots, a neighboring non-zero data element could be misidentified as part of the pilot block, causing ambiguity at the receiver and leading to incorrect decoding. Although these guard elements slightly reduce the available index space, their impact on the number of index bits is negligible. For example, with $N=21$, $L_{\mathrm{P}}=2$ and $K_{\mathrm{P}}=1$, the number of data index bits is $\lfloor \log_2 \binom{19}{2} \rfloor = 7$ without guard elements and $\lfloor \log_2 \binom{17}{2} \rfloor = 7$ with them, showing that the guard elements do not reduce the number of index bits in this case.

\emph{2) Codebook Spreading of DM-SVC}

After sparse mapping, the sparse vector is spread by the codebook matrix to generate the transmit signal. The spreading operation can be expressed as
\begin{equation}
	\mathbf{x}
	=
	\sqrt{\alpha}
	\sum_{k = 1}^{K_{\mathrm{D}}}
	s_k \mathbf{g}_k
	+
	\sqrt{1 - \alpha}
	\sum_{k' = 1}^{K_{\mathrm{P}}}
	\sum_{l = 1}^{L_{\mathrm{P}}}
	s_{k',l} \mathbf{g}_{k',l},
	\label{eq:DM_SVC_spreading_power}
\end{equation}
where $\mathbf{x} \in \mathbb{C}^M$ is the spread signal, $\alpha \in (0,1)$ is the power allocation rate between data and pilot components. $s_k$ is the $k$-th non-zero data symbol with codeword $\mathbf{g}_k$, and $s_{k',l}$ is the $l$-th pilot symbol in the $k'$-th non-zero pilot block with codeword $\mathbf{g}_{k',l}$. The entries of $\mathbf{G}$ are drawn independently from a symmetric Bernoulli distribution over $\{-1, +1\}$. For $M=4$ and $N=6$, an example codebook is given by
\begin{equation}
	{\bf{G}} = \beta \left[ {\begin{array}{*{20}{c}}
			{ - 1}&1&{ - 1}&1&{ - 1}&1\\
			1&{ - 1}&{ - 1}&{ - 1}&1&{ - 1}\\
			{ - 1}&1&{ - 1}&1&1&{ - 1}\\
			1&{ - 1}&1&1&{ - 1}&1
	\end{array}} \right],
\end{equation}
where $\beta = \sqrt{3/2(K_{\mathrm D}+K_{\mathrm P}L_{\mathrm P})(M_{\text{S}}-1)}$ is a normalization factor that ensures unit average symbol energy.

From \eqref{eq:DM_SVC_spreading_power}, the spread signal $\mathbf{x}$ is a linear superposition of two components: an element-sparse data part and a block-sparse pilot part. The power allocation factor $\alpha$ allows flexible adjustment of the relative power levels of data and pilot components, enabling a tradeoff between data detection performance and PN estimation accuracy.

\vspace{-0pt}
\subsection{Transmission of the Spread Sequence}

This subsection describes the transmission of the spread sequence $\mathbf{x}$ over an OFDM system in the presence of multipath fading and PN. A single-input single-output system is considered. The spread signal $\mathbf{x} \in \mathbb{C}^{M}$ is mapped onto frequency-domain subcarriers and transmitted via OFDM. Each OFDM symbol occupies $M$ subcarriers with spacing $\Delta f = 1/T_{\rm S}$, where $T_{\rm S}$ is the useful symbol duration.

Specifically, each OFDM symbol consists of $M$ subcarriers with subcarrier spacing $\Delta f = 1/T_{\rm S}$, where $T_{\rm S}$ denotes the OFDM symbol duration. The corresponding time-domain signal is obtained by applying the IDFT followed by CP insertion, i.e.,
\begin{equation}
	\mathbf{u} = \mathbf{A}_{\mathrm{CP}} \mathbf{F}^{\mathrm{H}} \mathbf{x},
\end{equation}
where $\mathbf{A}_{\mathrm{CP}} \in \mathbb{R}^{(M+L_{\mathrm{CP}})\times M}$ denotes the CP insertion matrix with CP length $L_{\mathrm{CP}}$.

The channel is modeled as a static multipath channel with $P$ resolvable paths. Let $\{h_p\}_{p=0}^{P-1}$ and $\{l_p\}_{p=0}^{P-1}$ denote the complex gain and delay index of the $p$-th path, respectively. In addition, the PN is assumed to be present at the receiver and modeled as a multiplicative impairment in the time domain. The received time-domain signal is \mbox{given by}
\begin{equation}
	r_m = p_m \sum_{p=0}^{P-1} h_p u_{m-\ell_p} + w_m,
\end{equation}
where $p_m$ is the $m$-th PN sample, and $w_m \sim \mathcal{CN}(0,\sigma_w^2)$ is an AWGN with variance $\sigma_w^2$.

The PN process is modeled as a discrete-time phase evolution using a first-order autoregressive (AR) model. Specifically, each PN sample is expressed as
\begin{equation}
	p_n = e^{j\vartheta_n},
	\label{eq:pn_sample}
\end{equation}
where $\vartheta_n$ is the instantaneous phase. We adopt an AR(1) model for $\vartheta_n$, given by
\begin{equation}
	\vartheta_n = \rho \vartheta_{n-1} + \epsilon_n,
	\label{eq:ar1_theta}
\end{equation}
where $\rho =e^{-2\pi f_{\mathrm{3dB}}{T_s}} \in (0,1)$ controls the temporal correlation of the PN and $\epsilon_n \sim \mathcal{N}(0,\sigma_{\mathrm{PN}}^2)$ is the innovation term. Following the commonly used PN bandwidth parametrization, the innovation variance is set as $\sigma_{\mathrm{PN}}^2 = 4\pi f_{\mathrm{3dB}} \Delta t$, where $f_{\mathrm{3dB}}$ denotes the 3-dB bandwidth of the PN and $\Delta t$ is the sampling interval. Stacking the received samples $\{ {r_m}\} _{m = 1}^{M + {L_{{\text{cp}}}}}$ into a vector $\mathbf{r}_{\mathrm t} \in \mathbb{C}^{(M+L_{\mathrm{cp}})}$, the received signal can be written in vector form as
\begin{equation}
	\mathbf{r}_{\mathrm t}
	=
	\operatorname{diag}(\mathbf{p})
	\mathbf{H}_{\mathrm{T}}
	\mathbf{A}_{\mathrm{cp}}
	\mathbf{F}^{\mathrm{H}}
	\mathbf{x}
	+
	\mathbf{w},
	\label{eq:td_rx_vector}
\end{equation}
where $\mathbf{p} \in \mathbb{C}^{(M+L_{\mathrm{CP}})}$ is the PN vector,  $\mathbf{H}_{\mathrm{T}} \in \mathbb{C}^{(M+L_{\mathrm{CP}})\times(M+L_{\mathrm{CP}})}$ is the time-domain channel matrix and $\mathbf{w} \sim \mathcal{CN}(\mathbf{0},\sigma_w^2\mathbf{I})$ is the noise vector. After CP removal and DFT processing, the frequency-domain received signal is
\begin{equation}
	\mathbf{y}
	=
	\mathbf{F}
	\mathbf{B}_{\mathrm{CP}}
	\mathbf{r}_{\mathrm t},
	\label{eq:fd_rx_def}
\end{equation}
where $\mathbf{B}_{\mathrm{CP}} \in \mathbb{R}^{M \times (M+L_{\mathrm{CP}})}$ is the CP removal matrix. Substituting \eqref{eq:td_rx_vector} into \eqref{eq:fd_rx_def} yields
\begin{equation}
	\mathbf{y}
	=
	\mathbf{F}
	\operatorname{diag}(\mathbf{p})
	\sum_{p=0}^{P-1}
	h_p
	\mathbf{T}^{p}
	\mathbf{F}^{\mathrm{H}}
	\mathbf{x}
	+
	\mathbf{\bar w},
	\label{eq:fd_rx_final}
\end{equation}
where ${\bf{\bar w}} = {\bf{Fw}}$, and $\mathbf{T} \in \mathbb{R}^{M \times M}$ denotes a permutation matrix given by
\begin{equation}
	\mathbf{T}
	=
	\begin{bmatrix}
		0      & 0      & \cdots & 0      & 1 \\
		1      & 0      & \cdots & 0      & 0 \\
		0      & 1      & \cdots & 0      & 0 \\
		\vdots & \vdots & \ddots & \vdots & \vdots \\
		0      & 0      & \cdots & 1      & 0
	\end{bmatrix}_{M\times M}.
	\label{eq:T_matrix}
\end{equation}

\subsection{BEM based PN Modeling}

In OFDM systems, PN introduces a multiplicative distortion in time domain, which translates into subcarrier-dependent impairments in the frequency domain. Since different subcarriers experience different effective phase rotations, PN estimation becomes particularly challenging in wideband systems. If PN is treated as an arbitrary unknown process across subcarriers, accurate estimation would require allocating an entire OFDM symbol as pilots, since estimating $M$ PN samples entails solving a system with $M$ unknowns. This overhead is prohibitive for short-packet transmission.

To address this issue, the BEM is used to model PN. By exploiting the inherent temporal correlation of PN, the BEM approximates the PN sequence using a small number of basis functions and corresponding coefficients. As a result, the PN estimation problem is transformed from estimating a large number of unknown PN samples to estimating a small number of BEM coefficients. This significantly reduces the required pilot overhead while preserving sufficient modeling accuracy. Therefore, BEM-based modeling provides an effective and practical approach for PN estimation in the considered DM-SVC framework.

We assume that perfect channel state information (CSI) is available at the receiver. This assumption is reasonable for two reasons. First, existing SVC studies have shown that sparse recovery-based decoding can achieve reliable data detection in static or quasi-static channels even without explicit CSI \cite{Ji2019}. Second, by assuming perfect CSI, the impact of PN on the reliability of SVC can be isolated and examined without being confounded by channel estimation errors. This assumption allows us to focus on the effect of PN as a single impairment and to gain clearer insights into the robustness of DM-SVC against PN.

We adopt the Slepian BEM for PN modeling. This choice is motivated by its optimal energy concentration property for bandlimited PN processes, which leads to faster eigenvalue decay and a more compact representation than conventional BEMs \cite{ZYFTWC25}. The PN vector can be represented as a linear combination of $Q$ ($Q \ll M$) Slepian basis functions and their corresponding coefficients, i.e.,
\begin{equation}
	{\bf{p}} = \sum\nolimits_{q = 0}^{Q - 1} {{c_q}{{\bf{b}}_q}}  + {\bf{w}_{\mathrm{mod}}},
\end{equation}
where $c_q$ is the $q$-th BEM coefficient, ${\mathbf{b}}_q$ is the corresponding Slepian basis vector, and ${\bf{w}_{\mathrm{mod}}}$ is the BEM modeling error. The Slepian basis functions are defined as the real-valued solutions to the eigenvalue problem:
\begin{equation}
	\sum_{n=0}^{M-1}
	\frac{\sin\!\big(2\pi f_{\max}(n-m)\big)}{\pi(n-m)}\, b_q[n]
	=
	\lambda_q b_q[m],
	\label{eq:slepian_eig}
\end{equation}
for $m=0,1,\ldots,M-1$, where $\lambda_q$ is the $q$-th eigenvalue, $b_q[m]$ denotes the $m$-th element of $\mathbf{b}_q$, and $f_{\max}$ is set according to the effective bandwidth of the PN process, while the number of basis functions $Q$ is chosen to balance modeling accuracy against estimation overhead. Thanks to the energy concentration property, a small number of basis functions suffices to capture the dominant PN dynamics. Each basis vector is normalized so that $\sum_{m=0}^{M-1} |b_q[m]|^2=1$.

Substituting (17) into (15) yields
\begin{equation}
{\bf{y}} = \sum\limits_{q = 0}^{Q - 1} {{c_q}} \;{\bf{F}}{\mathop{\rm diag}\nolimits} ({{\bf{b}}_q})\sum\limits_{p = 0}^{P-1} {{h_p}} {{\bf{T}}^{{p}}}{{\bf{F}}^{\rm{H}}}{\bf{x}} + {\bf{z}}.
\label{eq19}
\end{equation}
where $\mathbf{z}=\mathbf{F}\operatorname{diag}(\mathbf{w}_{\mathrm{mod}})
\sum_{p=0}^{P-1}h_p\mathbf{T}^{p}\mathbf{F}^{\mathrm H}\mathbf{x}
+\bar{\mathbf w}$ denotes the equivalent noise term, including the BEM modeling error and the additive noise.
With the BEM modeling, the number of PN-related parameters is reduced from $M$ to $Q$. However, the decoding remains challenging due to two factors. First, the pilot block positions are not predetermined, and the receiver must identify them before the pilot symbols become available for PN estimation. Second, strong mutual interference exists between data symbols and pilot symbols owing to their superposition in the spread domain. As a result, treating PN estimation and data decoding separately is ineffective.

To address the above challenges, an iterative joint PN estimation and data decoding algorithm is developed for the proposed DM-SVC scheme. The details of the proposed decoder is presented in Section~III.

\vspace{-0pt}
\section{Iterative Joint BEM PN Estimation and \\ Data Decoding}
In this section, an iterative joint PN estimation and data decoding algorithm is proposed for DM-SVC scheme. The decoding procedure is illustrated in Fig.~2. Because of the dual-sparse structure, PN estimation and data decoding are inherently coupled and cannot be performed separately. Specifically, the decoder first employs the CBMP algorithm to detect the $K_{\mathrm{P}}$ pilot blocks. Based on the estimated pilot block indices $\mathcal {\hat B}$ and the known pilot symbols $\mathbf{s}_{\mathrm{P}}$, a coarse PN estimate is obtained. The MMP algorithm is then applied to identify the indices of the $K_{\mathrm{D}}$ isolated non-zero data elements, followed by QAM symbol demodulation to recover the data information. The above PN estimation and data decoding procedures are performed in an iterative manner. In each iteration, the data symbols decoded in the previous iteration are treated as pseudo pilot symbols to enhance the PN estimation. By iteratively refining the PN estimate and decoded data, the proposed iterative algorithm achieves reliable PN estimation and data decoding. Finally, the detected pilot block indices are demapped to recover the $b_{\mathrm{P,I}}$ additional information bits carried by the block index pattern.

\vspace{-0pt}
\subsection{Pilot Block Detection}

According to the received frequency-domain signal model in \eqref{eq:fd_rx_final}, the received signal $\mathbf{y}$ can be rewritten as
\begin{equation}
	\mathbf{y}
	=
	\sqrt{1-\alpha}\,\mathbf{\Phi}\mathbf{s}_{\mathrm{P}}
	+
	\sqrt{\alpha}\,\mathbf{\Phi}\mathbf{s}_{\mathrm{D}}
	+
	\bar{\mathbf{w}},
	\label{eq:20}
\end{equation}
where $\mathbf{s}_{\mathrm{P}}$ and $\mathbf{s}_{\mathrm{D}}$ denote the pilot and data components of the sparse vector, respectively. $\bar{\mathbf{w}}$ is the effective noise comprising AWGN and modeling error, and the effective sensing matrix $\mathbf{\Phi}$ is given by
\begin{equation}
	\mathbf{\Phi}
	=
	\mathbf{F}
	\operatorname{diag}(\mathbf{p})
	\sum_{p=0}^{P-1}
	h_p
	\mathbf{T}^{p}
	\mathbf{F}^{\mathrm{H}}
	\mathbf G.
	\label{eq:Phi_def}
\end{equation}

For pilot block detection, the data component $\sqrt{\alpha}\,\mathbf{\Phi}\mathbf{s}_{\mathrm{D}}$ is treated as noise. Accordingly, the received signal can be approximated as
\begin{equation}
	\mathbf{y}
	=
	\sqrt{1-\alpha}\,\mathbf{\Phi}\mathbf{s}_{\mathrm{P}}
	+
	\tilde{\mathbf{w}},
	\label{eq:21}
\end{equation}
where $\tilde{\mathbf{w}}\triangleq\sqrt{\alpha}\,\mathbf{\Phi}\mathbf{s}_{\mathrm{D}}+\bar{\mathbf{w}}$.

Since $\mathbf{s}_{\mathrm{P}}$ has $K_{\mathrm{P}}$ non-zero blocks of length $L_{\mathrm{P}}$, the problem in \eqref{eq:21} can be formulated as a non-uniform block-sparse recovery problem
\begin{equation}
\begin{gathered}
	\min {\text{ }}\left\| {{\mathbf{y}} - \sqrt {1 - \alpha } {\mkern 1mu} {\mathbf{\Phi }}_{\mathcal{B}}{{\mathbf{s}}_\mathcal{B}}} \right\|_2^2\quad  \hfill \\
	{\text{s}}{\text{.t}}{\text{.}\quad|}\mathcal{B}{|} = {K_{\text{P}}}{L_{\text{P}}} \hfill \\ 
\end{gathered} .
	\label{eq:22}
\end{equation}

\begin{figure*}[htbp]
	\centerline{\includegraphics[width=0.85\textwidth]{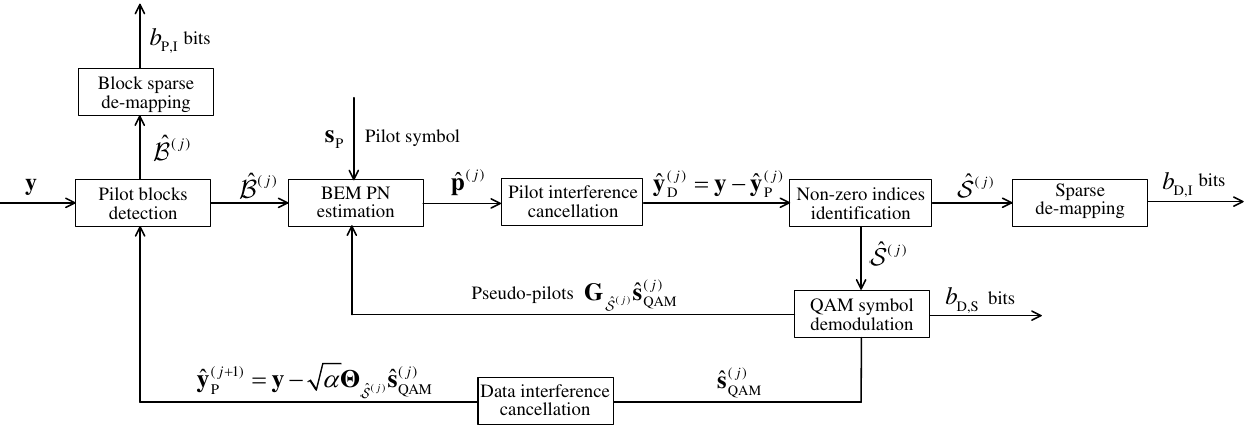}}
	\caption{Diagram of the iterative joint PN estimation and data decoding algorithm.}
	\label{fig2}
\vspace{-10pt}
\end{figure*}

\renewcommand{\algorithmicrequire}{\textbf{Input:}}
\renewcommand{\algorithmicensure}{\textbf{Output:}}
\begin{algorithm}[htbp]
	\caption{ {\bf {Algorithm 1~}}CBMP Algorithm for Pilot Block Detection }
	\begin{algorithmic}[1]
		\Require 
		received signal ${\bf{y}}$, measurement matrix $\bf \Phi$, number of non-zero blocks $K_{\mathrm P}$, length of non-zero block $L_{\mathrm P}$, power allocation ratio $\alpha$.
		\State Initialization: residual ${\bf r}^{(0)} = \bf y$, support set $ \mathcal{\hat B}_l^{(0)} = \emptyset $ for $l=1,2,\cdots,L_{\mathrm P}$, and $\mathcal{P} = \emptyset $.
		\While {$k<K_{\mathrm{P}}$}
		\For{$l=0,1,\cdots,L_{\mathrm P}-1$}
		\State Choose the block index $i^{(k)}$ that satisfies \eqref{eq26}.
		\State Combine support set: $ \mathcal{\hat B}_l^{(k)} = \mathcal{\hat B}_l^{(k - 1)} \cup \mathcal{P}$.
		\State Calculate ${\mathbf{\hat s}}^{(k)}$ according to \eqref{eq27}.
		\State Update residual according to \eqref{eq28}.
		\EndFor
		\EndWhile
		\State Find the optimal candidate set ${ \mathcal{\hat B}}$ from the $L_{\mathrm P}$ candidate sets following \eqref{eq29}. 
		\Ensure The estimated index set ${ \mathcal{\hat B}}$. 
	\end{algorithmic}
\end{algorithm}

Unlike standard block-sparse recovery, the $K_{\mathrm{P}}$ pilot blocks in DM-SVC are not aligned to a uniform block grid, and each block may start at an arbitrary position. The standard block orthogonal matching pursuit algorithm cannot be directly applied to solve \eqref{eq:22}.

To address this issue, the CBMP algorithm is proposed to solve \eqref{eq:22}. Similar to the first stage of the two-stage decoder in \cite{ZhangDMSVC2025}, the proposed CBMP algorithm exploits block correlations to identify non-zero pilot blocks. The core idea of CBMP is to generate candidate block supports by cyclically shifting a block window of length $L_{\mathrm{P}}$. According to the circular shift property of matrix-vector operations, one can obtain
\begin{equation}
	{\mathbf{y}} = \sqrt {1 - \alpha } {\mathbf{\Phi }}{{\mathbf{s}}_{\text{P}}} + {\mathbf{\tilde w}} = \sqrt {1 - \alpha } {\mathbf{\Phi }}{{\mathbf{\Pi }}^l}{{\mathbf{\Pi }}^{ - l}}{{\mathbf{s}}_{\text{P}}} + {\mathbf{\tilde w}},
	\label{eq:23}
\end{equation}
where ${\bf{\Pi }} = {\mathop{\rm circ}\nolimits} \{ [0,1,0, \cdots ,0]_{N \times 1}^{\rm{T}}\} $ is a permutation matrix. Eq. \eqref{eq:23} shows that the model is invariant under simultaneous circular shifts of ${{\bf{s}}_{\mathrm{P}}}$ and the columns of measurement matrix by $l$ positions. Exploiting this property, we can cyclically shift the vector ${{\bf{s}}_{\mathrm{P}}}$ up to ${L_{\mathrm{P}}} - 1$ times to ensure that at least one non-zero block aligns with the uniformly distributed block positions.

The CBMP algorithm first performs ${L_{\mathrm{P}}}$ cyclic shifts on the columns of ${\mathbf{\Phi }}$, and utilizes these new measurement matrices to construct ${L_{\mathrm{P}}}$ sub-problems:
\begin{equation}
{{ \mathcal{\hat B}}_l} = \mathop {\arg \min }\limits_{{\mathcal{B}_l}} \left\| {{\mathbf{y}} - \sqrt {1 - \alpha } {\mkern 1mu} {\mathbf{\Phi }}{{\mathbf{\Pi }}^l}{{\mathbf{s}}_{{\mathcal{B}_l}}}} \right\|_2^2
\label{eq:24}
\end{equation}
with $l = 0,1, \cdots ,{L_{\text{P}}} - 1$. 
For each sub-problem, the residual is initialized as ${{\mathbf{r}}^{(0)}} = {\mathbf{y}}$. At the $k$-th iteration, the block best correlated with ${{\mathbf{r}}^{(k - 1)}}$ is selected according to
\begin{equation}
	{i^{(k)}} = \mathop {\arg \max }\limits_i ||{\mathbf{\Psi }}_{l,i}^{\text{H}}{{\mathbf{r}}^{(k - 1)}}||_2^2,
	\label{eq26}
\end{equation}
where ${\mathbf{\Psi }}_{l} = \sqrt {1 - \alpha } {\mkern 1mu} {\mathbf{\Phi }}{{\mathbf{\Pi }}^l}$, and ${{\mathbf{\Psi }}_{l,i}} \in {\mathbb{R}^{M \times {{{L_{\text{P}}}}}}}$ is the $i$-th sub-matrix of ${{\mathbf{\Psi }}_l}$. The support set of $\mathbf s$ is updated as $ \mathcal{\hat B}_l^{(k)} =  \mathcal{\hat B}_l^{(k - 1)} \cup \mathcal{P}$, where $\mathcal{P} = \{ ({i^{(k)}} - 1)N/{L_{\text{P}}} + 1,({i^{(k)}} - 1)N/{L_{\text{P}}} + 2, \cdots ,{i^{(k)}}N/{L_{\text{P}}}\} $ is the set of chosen indices ${i^{(k)}}$. Once the support set is chosen, we find ${\mathbf{\hat s}}^{(k)}$ as the solution to
\begin{equation}
{{{\mathbf{\hat s}}}^{(k)}} = \mathop {\arg \min }\limits_{\mathcal{D} \subset  \mathcal{\hat B}_l^{(k)}} ||{\mathbf{y}} - {{\mathbf{\Psi }}_l}{{\mathbf{s}}_\mathcal{D}}||_2^2.
\label{eq27}
\end{equation}
The residual is then updated as
\begin{equation}
	{{\mathbf{r}}^{(k)}} = {\mathbf{y}} - \sum\nolimits_{i \in \mathcal{P}} {{{\mathbf{\Psi }}_{l,i}}} {{{\mathbf{\hat s}}}_i}.
	\label{eq28}
\end{equation}

After $K_{\mathrm{P}}$ iterations, the candidate set ${{ \mathcal{\hat B}}_l}$ of pilot blocks for the $l$-th cyclic shift is obtained. Finally, $L_{\mathrm{P}}$ candidate sets ${{\mathcal{\hat B}}_0},{{\mathcal{\hat B}}_1}, \cdots ,{{ \mathcal{\hat B}}_{{L_{\text{P}}} - 1}}$ are generated. The final pilot block support is chosen as the candidate that minimizes the residual:
\begin{equation}
	\mathcal{\hat B} = \mathop {\arg \min }\limits_{\mathcal{B} \subset \{ {{ \mathcal{\hat B}}_0},{{\mathcal{\hat B}}_1}, \cdots ,{{ \mathcal{\hat B}}_{{L_{\text{P}}} - 1}}\} } \left\| {{\mathbf{y}} - {{\mathbf{\Psi }}_l}{{\mathbf{s}}_\mathcal{B}}} \right\|_2^2.
	\label{eq29}
\end{equation}

The CBMP algorithm is summarized in \textbf{Algorithm 1}. The detected pilot block index ${ \mathcal{\hat B}}$ will be used for PN estimation, as described next.

\subsection{Coarse Estimation of PN}

Given the detected pilot block index set $\hat{\mathcal{B}}$ and the known pilot symbols ${{\mathbf{s}}_{{\text{nz}}}}$, the received signal model in \eqref{eq19} can be rewritten as
\begin{equation}
{\mathbf{y}} = {{\mathbf{A}}_{\text{P}}}{\mathbf{c}} + {{\mathbf{y}}_{\text{D}}} + {\mathbf{z}},
	\label{eq30}
\end{equation}
where ${{\mathbf{A}}_{\text{P}}} = [{[{{\mathbf{A}}_{\text{P}}}]_0},{[{{\mathbf{A}}_{\text{P}}}]_1}, \cdots ,{[{{\mathbf{A}}_{\text{P}}}]_{Q - 1}}] \in {\mathbb{C}^{M \times Q}}$ with its $q$-th column given by
\begin{equation}
{[{{\mathbf{A}}_{\text{P}}}]_q} = \sum\limits_{p = 0}^{P-1} {\sqrt {1 - \alpha } {\mathbf{F}}{{\mathbf{T}}^p}\operatorname{diag} ({{\mathbf{b}}_q} \odot {{\mathbf{h}}_p}){{\mathbf{F}}^{\text{H}}}{\mathbf{G}}{{[{{\mathbf{s}}_{\text{nz}}}]}_{ \mathcal{\hat B}}}}  ,
\label{eq31}
\end{equation}
where ${\mathbf{c}} = {[{c_0},{c_1}, \cdots ,{c_{Q - 1}}]^{\text{H}}}$ is the BEM coefficient vector, and ${\mathbf{y}}_{\mathrm{D}}$ denotes the received data signal.

The BEM coefficient vector $\mathbf{c}$ is then estimated using the least-squares (LS) method as
\begin{equation}
{\mathbf{\hat c}} = {({\mathbf{A}}_{\text{P}}^{\text{H}}{{\mathbf{A}}_{\text{P}}})^{ - 1}}{\mathbf{A}}_{\text{P}}^{\text{H}}{\mathbf{y}}.
	\label{eq32}
\end{equation}

The PN vector is then reconstructed as
\begin{equation}
	\hat{\mathbf{p}}
	=
	\mathbf{B}\hat{\mathbf{c}}.
	\label{eq33}
\end{equation}

The obtained $\hat{\mathbf{p}}$ provides a coarse estimate of the PN and will be further refined through the iterative joint PN estimation and data decoding procedure.

\renewcommand{\algorithmicrequire}{\textbf{Input:}}
\renewcommand{\algorithmicensure}{\textbf{Output:}}
\begin{algorithm}[htbp]
	\caption{ {\bf {Algorithm 2~}}Iterative PN Estimation and Data Decoding}
	\begin{algorithmic}[1]
		\Require 
		received signal ${\bf{y}}$, measurement matrix $\bf \Phi$, number of non-zero blocks $K$, length of non-zero block $L$.
        \State \textbf{Initialization Stage}: Detect the initial index set ${{ \mathcal{\hat B}}^{(0)}}$ of pilot blocks using the CBMP algorithm; Obtain the initial PN estimate ${{{\mathbf{\hat p}}}^{(0)}}$ according to Eqs. \eqref{eq32}-\eqref{eq33}.
        \State {\textbf{Iterative Stage}:}
		\While {$j<N_{\mathrm{iter}}$ and $\frac{{||{{{\bf{\hat p}}}^{(j + 1)}} - {{{\bf{\hat p}}}^{(j)}}|{|^2}}}{{||{{{\bf{\hat p}}}^{(j)}}|{|^2}}} < \varepsilon $}
	    \State Eliminate pilot interference based on \eqref{eq34} and \eqref{eq35}.
		\State Identify the indices of $K_{\mathrm D}$ non-zero elements by 
        \Statex \hspace{1.2 em} solving \eqref{eq36} using the MMP algorithm.
		\State Demodulate the $K_{\mathrm D}$ QAM symbols according to  \eqref{eq37}.        
		\State Update the measurement matrix based on Eq. \eqref{eq38}.
		\State Data interference cancellation according to \eqref{eq39}.
		\State Detect the pilot block indices using CBMP algorithm. 
		\State Estimate PN per \eqref{eq40}. 
		\EndWhile
		\Ensure Pilot blocks indices ${{ \mathcal{\hat B}}^*}$, non-zero data indices ${{ \mathcal{\hat S}}^*}$, estimated PN ${{{\mathbf{\hat p}}}^*}$, and demodulated QAM symbols ${\mathbf{\hat s}}_{{\text{QAM}}}^*$. 
	\end{algorithmic}
\end{algorithm}

\subsection{Data Decoding}

In the following, the superscript ${( \cdot )^{(j)}}$ denotes quantities at the $j$-th iteration. With the initial pilot block support $\hat{\mathcal{B}}^{(0)}$ and the estimated PN vector $\hat{\mathbf{p}}^{(0)}$ in hand, the data decoding proceeds in two steps. The pilot-induced component is first subtracted from the received signal, and the remaining signal is then used to detect the data indices and demodulate the QAM symbols.

At the $j$-th iteration ($j>0$), the estimated pilot-related received signal can be reconstructed as
\begin{equation}
{{\mathbf{\hat y}}_{\text{P}}^{(j)}} = \sqrt {1 - \alpha } {\mkern 1mu} {{\mathbf{\Phi }}_{ \mathcal{\hat B}^{(j-1)}}}{[{{\mathbf{s}}_{\text{P}}}]_{ \mathcal{\hat B}^{(j-1)}}},
	\label{eq34}
\end{equation}
where ${{\mathbf{\Phi }}_{ \mathcal{\hat B}^{(j-1)}}} = {\mathbf{F}}\operatorname{diag} ({\mathbf{\hat p}^{(j-1)}})\sum\limits_{p = 0}^{P - 1} {{h_p}{{\mathbf{T}}^p}{{\mathbf{F}}^{\text{H}}}} {{\mathbf{G}}_{ \mathcal{\hat B}^{(j-1)}}}$.

By removing the pilot signal in \eqref{eq34} from the received signal $\mathbf{y}$, the data-dominated received signal is obtained as
\begin{equation}
	\mathbf{\hat y}_{\mathrm{D}}^{(j)}
	=
	\mathbf{y}
	-
	\mathbf{\hat y}_{\mathrm{P}}^{(j)}.
	\label{eq35}
\end{equation}

Based on \eqref{eq35}, the support detection problem can be formulated as the following sparse recovery problem:
\begin{equation}
\{  \mathcal{\hat S}^{(j)},{[{{{\mathbf{\hat s}}}_{\text{D}}}]_{ \mathcal{\hat S}^{(j)}}}\}  = \mathop {\arg \min }\limits_{|\mathcal{S}| = {K_{\text{D}}}} \left\| {{{\mathbf{y}}_{\text{D}}^{(j)}} - \sqrt \alpha  {\mkern 1mu} {\mkern 1mu} {{\mathbf{\Phi }}_\mathcal{S}}{{[{{\mathbf{s}}_{\text{D}}}]}_\mathcal{S}}} \right\|_2^2.
	\label{eq36}
\end{equation}

In this paper, the sparse recovery problem in \eqref{eq36} is solved using the MMP algorithm \cite{Ji2018}.The MMP algorithm iteratively explores multiple candidate support paths to mitigate error propagation and selects the support set that minimizes the residual. Then, the estimated data support set $\hat{\mathcal{S}}^{(j)}$ and the corresponding non-zero values ${[{{{\mathbf{\hat s}}}_{\text{D}}}]_{ \mathcal{\hat S}^{(j)}}}$ are obtained.

Given the estimated non-zero values, the demodulated symbol $\hat{a}_k$ corresponding to the $k$-th detected non-zero element can be obtained as
\begin{equation}
{{\hat s}_{{\text{QAM,}}k}^{(j)}} = \mathop {\arg \min }\limits_{{a_i} \in \mathcal{A}} {\left| {{{[{{\hat s}_{\text{D}}}]}_{{{ \mathcal{\hat S}}_k}^{(j)}}} - {a_i}} \right|^2},
	\label{eq37}
\end{equation}
where $\mathcal{A} = \{ {a_1},{a_2}, \cdots ,{a_{{M_{\mathrm S}}}}\} $ is the ${M_{\mathrm S }}$-ary alphabet. 

\subsection{Iterative Joint PN Estimation and Data Decoding}
The iterative joint PN estimation and data decoding algorithm is summarized in \textbf{Algorithm 2}. The algorithm procedure is detailed as follows. The CBMP algorithm detects the pilot block support, and a coarse PN estimate is obtained from the known pilot symbols as described in Sections III-A and III-B. In each subsequent iteration, the receiver subtracts the reconstructed pilot component from the received signal and applies the MMP algorithm to detect the data support and demodulate the QAM symbols. The demodulated data symbols are then incorporated as pseudo-pilots to update the measurement matrix. At the $(j+1)$-th iteration, the $q$-th column of the updated matrix is given by
\begin{equation}
[{{\mathbf{A}}_{\text{P}}}]_q^{(j+1)} = \sqrt \alpha  {{\mathbf{\Theta }}_{{{ \mathcal{\hat S}}^{(j)}}}}{\mathbf{\hat s}}_{{\text{QAM}}}^{(j)} + \sqrt {1 - \alpha } {{\mathbf{\Theta }}_{{{ \mathcal{\hat B}}^{(j)}}}}{[{{\mathbf{s}}_{\text{P}}}]_{{{ \mathcal{\hat B}}^{(j)}}}},
\label{eq38}
\end{equation}
where ${\mathbf{\Theta }} = \sum\nolimits_{p = 0}^{P - 1} {{\mathbf{F}}{{\mathbf{T}}^p}\operatorname{diag} ({{\mathbf{b}}_q} \odot {{\mathbf{h}}_p}){{\mathbf{F}}^{\text{H}}}{\mathbf{G}}} $, and ${\mathbf{\hat s}}_{{\text{QAM}}}^{(j)}$$ = {[\hat s_{{\text{QAM,1}}}^{(j)},\hat s_{{\text{QAM,2}}}^{(j)}, \cdots ,\hat s_{{\text{QAM,}}{K_{\text{D}}}}^{(j)}]^{\text{H}}} \in {\mathbb{R}^{{K_{\text{D}}}}}$. The measurement matrix ${\mathbf{A}}_{\text{P}}^{(j+1)} = [[{{\mathbf{A}}_{\text{P}}}]_0^{(j+1)},[{{\mathbf{A}}_{\text{P}}}]_1^{(j+1)}, \cdots ,[{{\mathbf{A}}_{\text{P}}}]_{Q - 1}^{(j+1)}]$ is used to refine both pilot block detection and PN estimation. The pilot-related received signal is estimated by subtracting the data contribution from $\mathbf y$, i.e.,
\begin{equation}
	{\mathbf{\hat y}}_{\text{P}}^{(j + 1)} = {\mathbf{y}} - \sqrt \alpha  {{\mathbf{\Theta }}_{{{ \mathcal{\hat S}}^{(j)}}}}{\mathbf{\hat s}}_{{\text{QAM}}}^{(j )}.
	\label{eq39}
\end{equation}

At the $(j+1)$-th iteration, the PN is updated by substituting ${\mathbf{\hat y}}_{\text{P}}^{(j + 1)}$ for $\mathbf y$ in the LS estimation procedure, yielding
\begin{equation}
	{{{\mathbf{\hat p}}}^{(j + 1)}} = {\mathbf{B}}{{{\mathbf{\hat c}}}^{(j + 1)}},
	\label{eq40}
\end{equation}
where ${{{\mathbf{\hat c}}}^{(j + 1)}} = {({({\mathbf{A}}_{\text{P}}^{(j + 1)})^{\text{H}}}{\mathbf{A}}_{\text{P}}^{(j + 1)})^{ - 1}}{({\mathbf{A}}_{\text{P}}^{(j + 1)})^{\text{H}}}{\mathbf{\hat y}}_{\text{P}}^{(j + 1)}$.

The iteration terminates when the relative change in the PN estimate falls below a threshold $\varepsilon$, i.e., $||{{{\mathbf{\hat p}}}^{(j + 1)}} - {{{\mathbf{\hat p}}}^{(j)}}|{|^2}/||{{{\mathbf{\hat p}}}^{(j)}}|{|^2} < \varepsilon $, or when a maximum number of iterations is reached. In our simulations, $\varepsilon = 10^{-5}$. Algorithm~2 outputs the final pilot block support ${{ \mathcal{\hat B}}^*}$, non-zero data indices ${{ \mathcal{\hat S}}^*}$, and demodulated QAM symbols ${\mathbf{\hat s}}_{{\text{QAM}}}^*$, from which the $b$ bits can be recovered.

\section{Performance Analysis}
In this section, the computational complexity, SE, and codebook storage overhead of the proposed DM-SVC scheme are analyzed.

\subsection{Complexity Analysis}
The complexity of the proposed iterative joint PN estimation and data decoding algorithm arises from three components, i.e., pilot block detection, PN estimation, and data decoding. The pilot block detection is performed using the CBMP algorithm. The complexity of computing block correlations in each circular shift is $\mathcal{O}(MN)$, and the complexity of identifying non-zero block indices is $O(ML_{\text{P}}^2 + L_{\text{P}}^3)$. Since the CBMP algorithm runs $K_{\mathrm P}$ iterations with $L_{\mathrm P}$ circular shifts per iteration, the total complexity can be summarized as $O\left( {{K_{\text{P}}}{L_{\text{P}}}(MN + ML_{\text{P}}^2 + L_{\text{P}}^3)} \right)$. PN estimation is dominated by the LS computation of the BEM coefficients, with complexity $O(M{Q^2} + {Q^3})$. The complexity of the data decoding part is mainly concentrated in the MMP algorithm, with complexity $O\left( {MK_{\text{D}}^2 + {\tilde L}K_{\text{D}}(MN + 2M)} \right)$, where $\tilde L$ is the maximum searching paths in the MMP \mbox{algorithm \cite{Ji2018}}. In practice short-packet transmission scenarios, $K$, $Q$, and $L_{\mathrm{P}}$ are much smaller than $M$ and $N$. Over $N_{\mathrm {iter}}$ iterations, the overall complexity of the proposed iterative decoding algorithm can be summarized as $O\left( {{N_{{\rm{iter}}}}\left( {MN({K_{\rm{P}}}{L_{\rm{P}}} + \tilde L{K_{\rm{D}}}) + M{K_{\rm{P}}}L_{\rm{P}}^3} \right)} \right)$.

If the conventional MMP decoder is directly applied to DM-SVC, the pilot and data blocks are jointly detected without exploiting the proposed two-stage structure. In this case, the number of non-zero elements becomes ${K_{\rm{D}}} + {K_{\rm{P}}}{L_{\rm{P}}}$. and the corresponding complexity can be expressed as $O\left( {M{{({K_{\rm{D}}} + {K_{\rm{P}}}{L_{\rm{P}}})}^2} + \tilde L({K_{\rm{D}}} + {K_{\rm{P}}}{L_{\rm{P}}})(MN + 2M)} \right)$.

The numerical complexity comparison between proposed two-stage decoder and the conventional MMP decoder is summarized in Table I. It can be observed that the proposed decoder achieves a complexity comparable to that of the MMP decoder. When the number of non-zero pilot elements increases, the complexity of the proposed decoder becomes slightly lower than that of the MMP decoder. This is because the proposed two-stage decoder separates pilot block detection from data decoding. Consequently, it avoids the enlarged joint search space encountered by MMP when all non-zero elements are jointly treated as data symbols.

\begin{table}[t]
\renewcommand{\arraystretch}{1.2}
\centering
\caption{Numerical complexity comparison of different decoders under various coding parameters, with $M=96$, $N=140$, $K_{\rm D}=2$ and $\tilde L=2$.}
\begin{tabular}{|c|c|c|}
\hline
Item & $K_{\rm P}=1$, $L_{\rm P}=2$ & $K_{\rm P}=2$, $L_{\rm P}=4$ \\ \hline
Proposed decoder & $2.4\times 10^{5}$ & $4.9\times 10^{5}$ \\ \hline
MMP decoder & $2.2\times 10^{5}$ & $6.5\times 10^{5}$ \\ \hline
\end{tabular}
\end{table}

\subsection{SE Analysis}
Each transmission block is assumed to occupy one OFDM symbol with ${T_{\rm{s}}}\Delta f = 1$, where ${T_{\rm{s}}}$ is the OFDM symbol duration and $\Delta f$ is the subcarrier spacing. Following the definitions in Section II, the SE of the conventional SSC scheme \cite{ZhangXuewan2022} can be expressed as
\begin{equation}
\mathrm{SE}_{\mathrm{SSC}} = \frac{b_{\mathrm{D},\mathrm{I}} + b_{\mathrm{S},\mathrm{I}}}{M} \quad \mathrm{(bps/Hz)}.
\end{equation}
In contrast, the SE of proposed DM-SVC scheme is given by
\begin{equation}
\mathrm{SE}_{\mathrm{DM-SVC}} = \frac{b_{\mathrm{P},\mathrm{I}} + b_{\mathrm{D},\mathrm{I}} + b_{\mathrm{S},\mathrm{I}}}{M} \quad \mathrm{(bps/Hz)}.
\end{equation}

\begin{figure}[htbp]
\centerline{\includegraphics[width=0.48\textwidth]{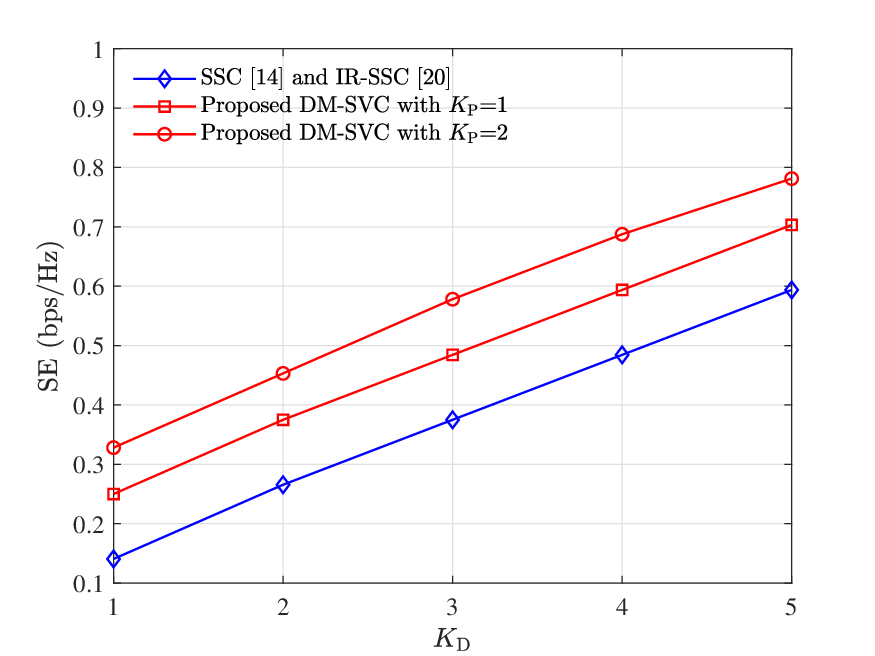}}
\caption{SE comparison between the proposed DM-SVC and existing SVC schemes with $L_{\mathrm P}=2$ and $M=72$.\vspace{-0pt}}
\end{figure}

Comparing (41) and (42) shows that DM-SVC achieves a higher SE than SSC by using the pilot block indices as an additional information-bearing dimension. Fig.~3 compares the SE of DM-SVC and SSC as a function of $K_{\rm{D}}$. It can be observed that the proposed DM-SVC scheme consistently achieves higher SE for different values of $K_{\rm{D}}$. For example, the DM-SVC scheme provides at least an 18\% SE gain over the conventional SSC scheme. Moreover, increasing the number of pilot blocks $K_{\rm{P}}$ further improves the SE of DM-SVC, since more pilot block indices become available to carry information bits.

\subsection{Codebook Storage Overhead Analysis}
The codebook storage overhead is measured by the number of real-valued entries to be stored, which scales as $\mathcal{O}(MN)$. Fig. 4 compares the codebook storage overhead of DM-SVC scheme with existing SSC \cite{ZhangXuewan2022} and IR-SSC \cite{Zhangxue25} schemes as a function of the number of transmitted bits $b$ per block. For conventional SSC and IR-SSC schemes, the codebook storage overhead increases exponentially with $b$. In contrast, the proposed DM-SVC scheme demonstrates substantially lower storage requirements. This is because DM-SVC encodes additional bits through the pilot block indices rather than extending the codebook size. For a given transmission rate, DM-SVC therefore requires less memory for codebook storage, making it attractive for storage-constrained terminals in short-packet systems.

\begin{figure}[htbp]
\centerline{\includegraphics[width=0.48\textwidth]{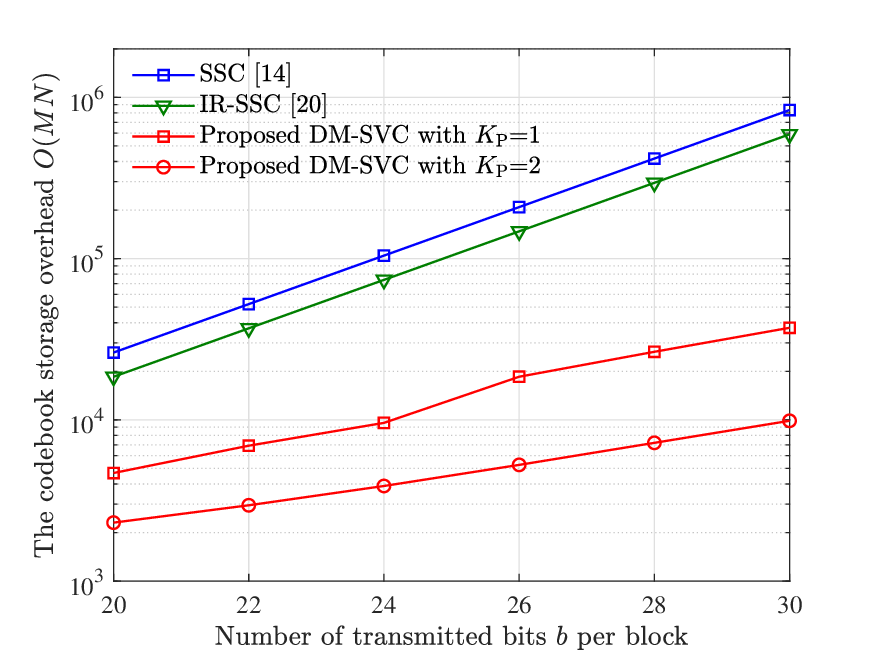}}
\caption{Codebook storage overhead comparison for different schemes with $K_{\mathrm D}=2$, $L_{\mathrm P}=2$ and $M=72$.\vspace{-0pt}}
\end{figure}

\vspace{-0pt}
\section{Simulation Results} 
In this section, Monte Carlo simulations are conducted to evaluate the BLER performance and PN estimation accuracy of the proposed DM-SVC scheme. The effects of several key parameters on BLER performance are also investigated, including the power allocation factor $\alpha$, pilot block length $L_{\mathrm P}$, BEM order $Q$, number of subcarriers $M$, and the PN strength. Unless otherwise specified, the default simulation parameters are listed in Table II. The BLER is defined as the ratio of erroneously received packets to the total number of transmitted packets. The normalized mean square error (NMSE) of the PN estimate is defined as
\begin{equation}
{\rm{NMSE}} = \mathbb{E}\{ {{||{\bf{p}} - {\bf{\hat p}}||_2^2} \over {||{\bf{p}}||_2^2}}\}. 
\end{equation}

\begin{table}
\renewcommand{\arraystretch}{1.2}
\centering
\caption{Simulation Parameters}
\begin{tabular}{|c|c|}
\hline
\textbf{System parameter} & \textbf{Value} \\
\hline
Carrier frequency ($f_{\rm c}$) & 3 GHz \\
\hline
Subcarrier spacing ($\Delta f$) & 30 KHz \\
\hline
Order of Slepian BEM ($Q$)& 2 \\
\hline
Modulation scheme & QPSK \\
\hline
The 3-dB bandwidth of PN & 1000 Hz \\
\hline
Channel model & 5G TDL-A \cite{5g-channel} \\
\hline
\end{tabular}
\label{tab3}
\end{table}

Three categories of benchmark schemes are considered for performance comparison.
\begin{itemize}
\item \textbf{Conventional SVC schemes with dedicated pilots}. The SSC \cite{ZhangXuewan2022}, GSPARCs \cite{MSinhaTCOM2024}, and index redefinition SSC (IR-SSC) \cite{Zhangxue25} schemes with dedicated pilots serve as benchmarks, denoted by SSC w/ DP, GSPARCs w/ DP, and IR-SSC w/ DP, respectively. In these schemes, the pilot and data parts are transmitted over two consecutive OFDM symbols. The first OFDM symbol is used for PN estimation, while the second OFDM symbol carries the data. Therefore, the numbers of pilot and data symbols are both equal to M. To provide a favorable benchmark for the DP schemes, it is assumed that the PN affecting the pilot symbols is the same as that affecting the data symbols. The PN is estimated from the pilot symbols via the linear minimum mean square error (LMMSE) algorithm, after which data decoding is performed based on the estimated PN. For data decoding, the SSC and IR-SSC schemes employ the MMP algorithm, while the GSPARCs adopts the match-and-decode algorithm as described \mbox{in \cite{MSinhaTCOM2024}.}
\item \textbf{Conventional SVC schemes with superimposed pilots}. The superimposed-pilot versions of SSC, GSPARCs, and IR-SSC schemes are also considered, denoted by SSC w/ SP, GSPARCs w/ SP, and IR-SSC w/ SP, respectively. The pilot symbols are superimposed onto the data symbols and transmitted over the same OFDM symbol. The PN is modeled using the BEM ($Q=2$), and its coefficients are estimated from the received signal. Let $\beta$ denote the power allocation ratio of data, the power allocation ratio of pilot is $1-\beta$. Similar to the proposed DM-SVC scheme, the power allocation between data and pilots is optimized through Monte Carlo simulations for each SNR point. The same decoding algorithms as in the dedicated-pilot case are adopted.
\item \textbf{DM-SVC scheme with perfect PN knowledge}. In this scheme, the receiver is assumed to have perfect knowledge of the PN. The CBMP algorithm is first used to detect the pilot block support, and the corresponding signal component is subtracted from the received signal. The MMP algorithm is then applied to detect the indices of the isolated non-zero data elements and demodulate the QAM symbols.
\end{itemize}

\subsection{BLER Performance}

Fig. 5 illustrates the impact of data power allocation ratio $\alpha$ on BLER under different SNR levels. The BLER exhibits a convex trend with respect to $\alpha$, indicating the existence of an optimal power allocation region. When $\alpha$ is too small, excessive power is assigned to pilot blocks. Although this improves the PN estimation accuracy, it leaves insufficient power for data symbols and thus degrades the BLER performance. When $\alpha$ is too large, more power is allocated to data symbols, but the reduced pilot power degrades the accuracy of PN estimation and consequently degrades the decoding reliability. As the SNR increases, the optimal value of $\alpha$ shifts slightly but remains within a relatively narrow range. This can be explained by the fact that the pilot power directly affects the accuracy of BEM-based PN estimation, while the resulting residual PN further influences sparse support recovery and symbol detection. Owing to this coupling between PN estimation and data decoding, the BLER is a highly non-linear function of $\alpha$, and thus a closed-form optimal $\alpha$ is difficult to obtain in the DM-SVC framework. From a practical perspective, the results suggest that $\alpha$ does not need to be precisely optimized for each SNR value. Instead, a fixed $\alpha$ selected from this near-optimal range can achieve stable BLER performance over a wide SNR range, thereby simplifying system design and reducing power control overhead.

\begin{figure}
\centerline{\includegraphics[width=0.46\textwidth]{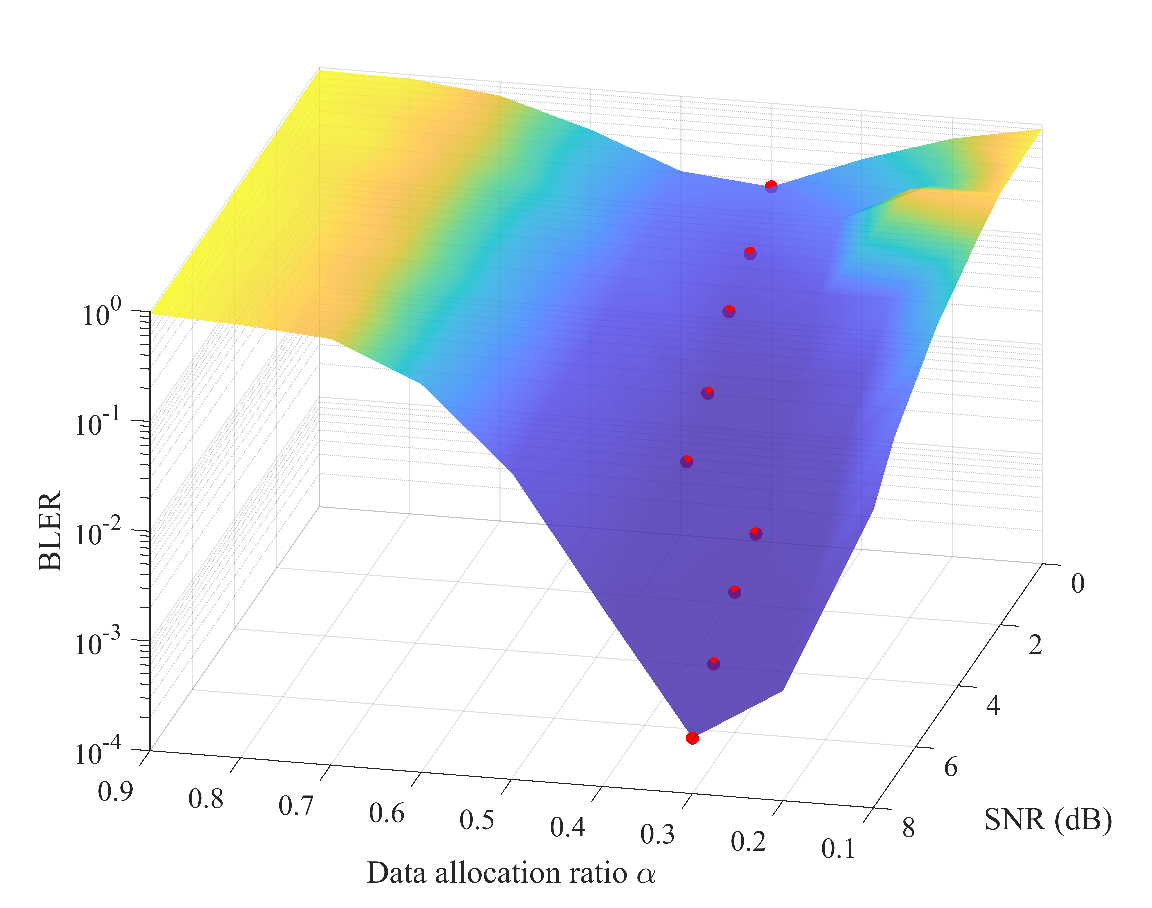}}\vspace{-6pt}
\caption{Impact of power allocation ratio $\alpha$ on BLER performance under different SNRs, with $M=72$, and $b=24$.\vspace{-5pt}}
\end{figure}
\begin{figure}[htbp]
\centerline{\includegraphics[width=0.46\textwidth]{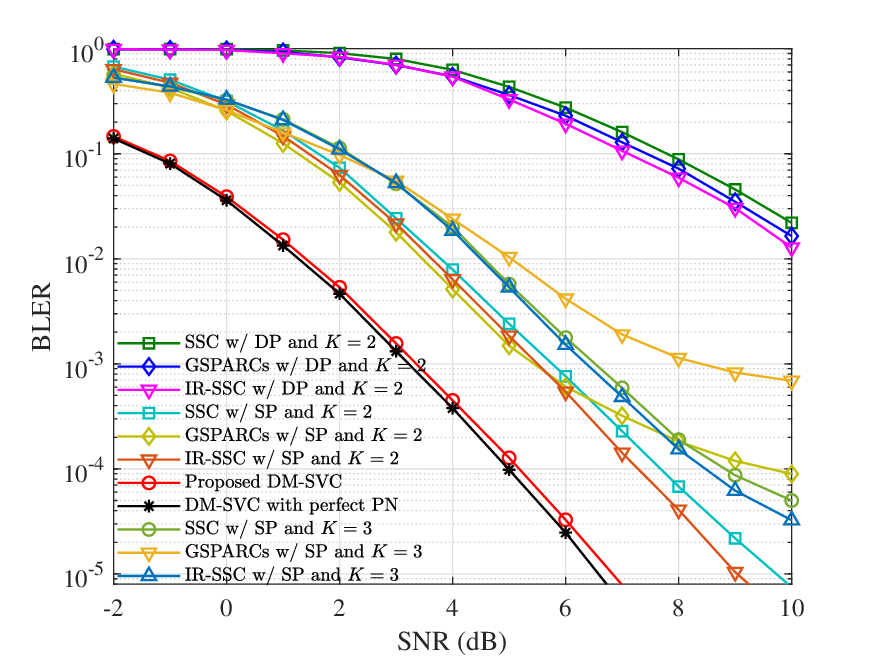}}
\caption{BLER performance comparison for different schemes with $K_{\mathrm D}=2$, $K_{\mathrm P}=1$, $L_{\mathrm P}=2$, $M=96$, and $b=24$.\vspace{-0pt}}
\end{figure}

Fig.~6 compares the BLER performance of the proposed DM-SVC scheme with benchmark schemes. The proposed DM-SVC scheme outperforms the existing schemes over the entire SNR range and achieves about a 2 dB SNR gain over IR-SSC at a BLER of $10^{-5}$. This gain mainly stems from the iterative interaction between PN estimation and sparse support recovery. More accurate BEM-based PN estimation suppresses phase distortion and improves support recovery, while the refined support detection further enhances the subsequent PN estimation, leading to decoding performance improvement. Conventional SVC schemes with dedicated pilots do not employ BEM-based PN modeling, and their PN estimation relies on noisy pilot observations, which leads to inaccurate PN compensation and degraded decoding reliability. For the superimposed pilot schemes, although the accuracy of PN estimation is improved, their pilots do not carry information bits. As a result, when $K=2$, to transmit the same number of bits, these schemes require significantly longer sparse vectors (e.g., $N=726$ for IR-SSC, $N=1450$ for SSC, and $N=2060$ for GSPARCs), compared to $N=132$ in DM-SVC scheme. The increased vector dimension leads to higher codebook correlation and a larger search space, which degrades sparse support recovery performance. When the sparsity level of conventional SVC schemes increases from $K=2$ to $K=3$, the required sparse vector length can be reduced. However, the increased number of non-zero elements also expands the support search space, and thus the BLER performance is still limited. In contrast, DM-SVC embeds information into pilot block indices, enabling a more compact sparse vector and improved detection reliability.

Moreover, although DM-SVC requires joint detection of pilot block indices and data symbols, the constant-modulus property of PN does not affect the identification of non-zero pilot block indices.
The pilot block detection is mainly influenced by noise and data interference, which can be effectively handled by sparse recovery algorithm. As a result, the performance gap between DM-SVC under practical PN and the perfect-PN baseline remains within 0.3 dB at a BLER of $1\times 10^{-5}$, demonstrating its robustness to PN.

\begin{figure}[htbp]
\centerline{\includegraphics[width=0.46\textwidth]{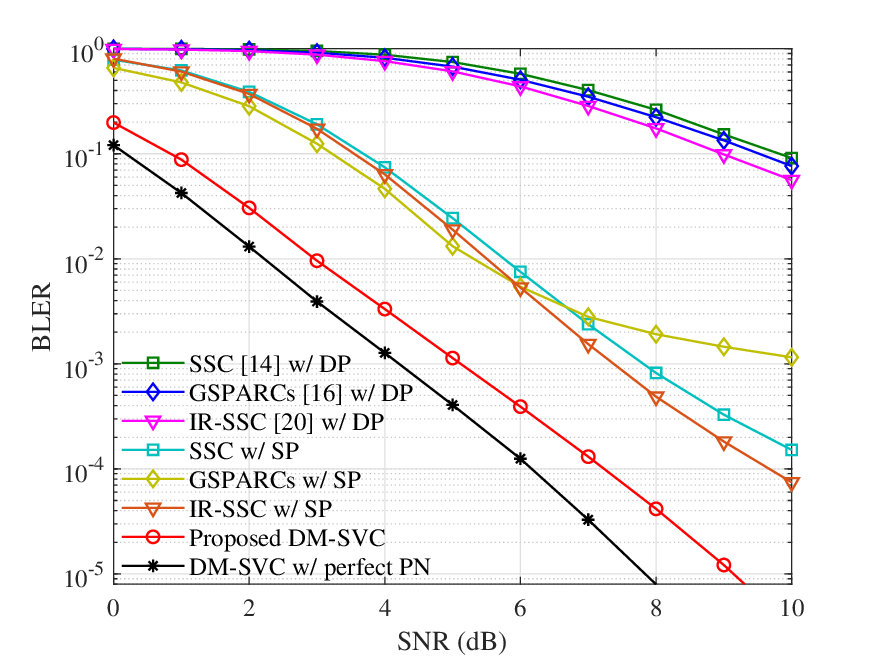}}
\caption{BLER performance comparison for different schemes with $K_{\mathrm D}=3$, $K_{\mathrm P}=1$, $L_{\mathrm P}=2$, $M=96$ and $b=31$.\vspace{-0pt}}
\end{figure}
\begin{figure}[htbp]
\centerline{\includegraphics[width=0.46\textwidth]{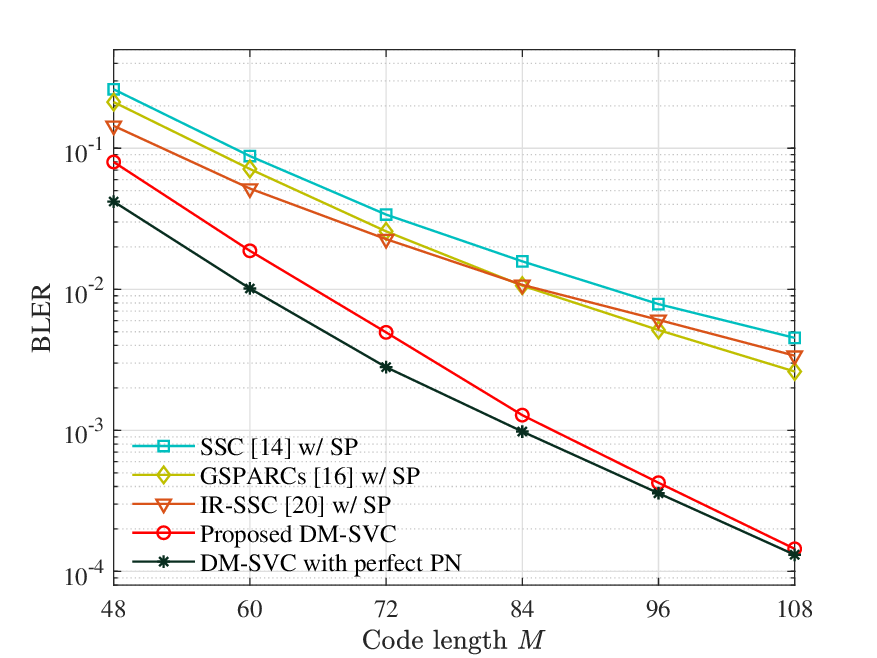}}
\caption{BLER of different schemes versus code length $M$ with $K_{\mathrm D}=2$, $K_{\mathrm P}=1$, $L_{\mathrm P}=2$, $b=24$ and SNR = 4 dB.\vspace{-10pt}}
\end{figure}

Fig.~7 shows similar trends for $K_{\mathrm D}=3$ and $b=31$. Even with a higher bit payload, the proposed DM-SVC still maintains a clear BLER advantage over the baseline schemes. This shows that the joint pilot and data design remains effective under more challenging detection conditions, since it can still provide reliable PN estimation while avoiding the large sparse-vector dimension required by the conventional SVC schemes.

\begin{figure}[htbp]
\centerline{\includegraphics[width=0.46\textwidth]{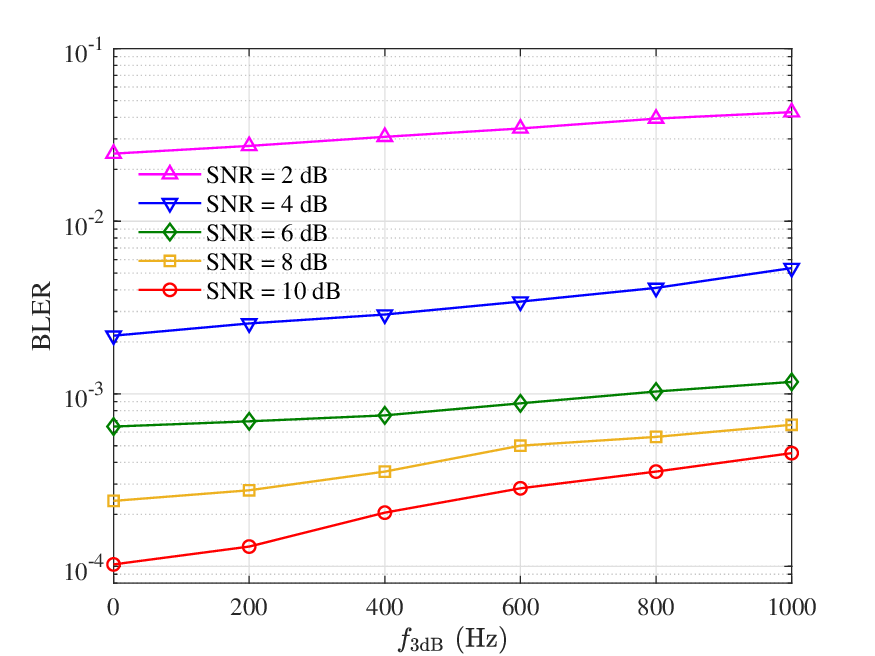}}\vspace{-0pt}
\caption{BLER as a function of $f_{\mathrm {3dB}}$ with $K_{\mathrm D}=2$, $K_{\mathrm P}=1$, $L_{\mathrm P}=2$, $M=72$ and $b=24$.\vspace{-15pt}}
\end{figure}
\begin{figure}[htbp]
\centerline{\includegraphics[width=0.46\textwidth]{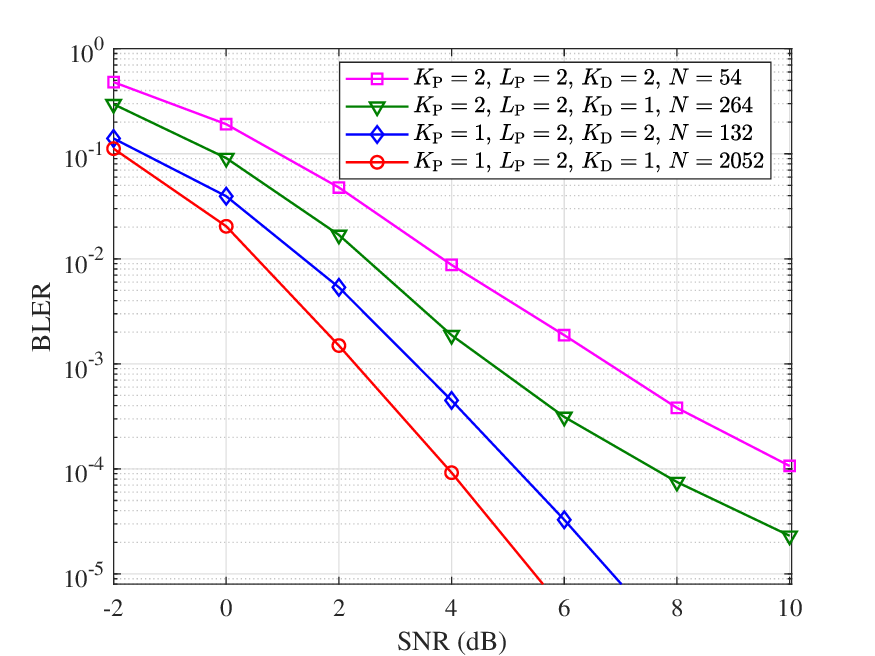}}
\caption{BLER performance of DM-SVC under different pilot and data sparsity configurations with $M=96$ and $b=24$.\vspace{-10pt}}
\end{figure}

Fig. 8 shows the BLER performance of different schemes as a function of the code length $M$. The proposed DM-SVC scheme achieves consistent BLER improvement as 
$M$ increases, whereas the benchmark schemes exhibit limited improvement. For example, when $M=96$, the BLER of DM-SVC is approximately one order of magnitude lower than that of the GSPARCs scheme. This is because DM-SVC can jointly exploit the increased observations for BEM-based PN estimation and data decoding as $M$ grows. In contrast, conventional schemes rely mainly on fixed pilots for PN estimation, thereby limiting the BLER improvement obtained from increasing $M$. Moreover, the gap between DM-SVC scheme under practical PN and the perfect-PN case decreases as $M$ increases, indicating that a longer block length provides more reliable PN estimation and reduces the residual phase distortion. This indicates that simply increasing the code length does not necessarily bring a substantial BLER improvement. Its benefit becomes significant when the additional observations can be effectively used to improve both BEM-based PN estimation and data decoding.

Fig. 9 shows the BLER performance of the proposed DM-SVC as a function of the PN 3-dB bandwidth at different SNRs. The results show that the BLER increases with $f_{\text{3dB}}$ at all considered SNR levels, because faster phase variations lead to larger accumulated phase distortion within each transmission block, making accurate symbol recovery more difficult. Nevertheless, the performance degradation remains limited even as $f_{\text{3dB}}$ increases to 1000 Hz. For instance, at an SNR of 10 dB, the BLER rises only from $10^{-4}$ to $5 \times 10^{-4}$. This result indicates that the iterative decoder effectively tracks and compensates for the phase rotations within each block. It is noteworthy that as $f_{\text{3dB}}$ increases, the temporal correlation of the PN process decreases, making it more difficult to accurately capture the PN process using a low-order BEM. This results in increased modeling errors and larger residual PN after PN estimation. This suggests that the BEM order should be properly increased with the PN 3-dB bandwidth to accurately capture fast phase variations and ensure reliable decoding.

\begin{figure}[htbp]
\centerline{\includegraphics[width=0.46\textwidth]{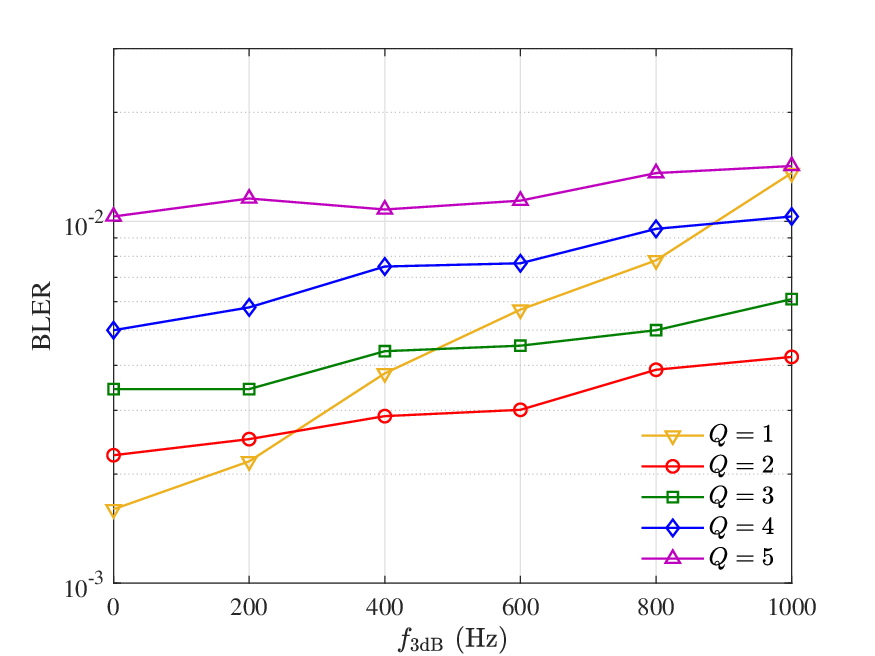}}
\caption{Impact of BEM order $Q$ on BLER performance under different PNs with $M=72$ and SNR = 4 dB.\vspace{-5pt}}
\end{figure}

Fig. 10 compares the BLER performance of the proposed DM-SVC scheme under different pilot and data sparsity configurations with a fixed information payload. It can be seen that increasing either $K_{\mathrm P}$ or $K_{\mathrm D}$ degrades BLER performance but substantially reduces the required sparse vector length $N$. Specifically, the configuration with $K_{\mathrm P}=1$ and $K_{\mathrm D}=1$ achieves the best BLER performance. However, this configuration requires the largest sparse vector length $N=2052$, resulting in the highest codebook size and decoding complexity. In contrast, the configuration with $K_{\mathrm P}=2$ and $K_{\mathrm D}=2$ requires only $N=54$, significantly reducing the codebook size, but exhibits the worst BLER across the entire SNR range. Meanwhile, the configurations with $K_{\mathrm P}=1$, $K_{\mathrm D}=2$, $N=132$, and $K_{\mathrm P}=2$, $K_{\mathrm D}=1$, $N=264$ offer intermediate trade-offs between reliability and complexity. This indicates that the DM-SVC scheme provides flexibility in balancing BLER performance and decoding complexity through appropriate encoding parameter selection.

Fig.~11 investigates the impact of the BEM order $Q$ on decoding reliability under different PN bandwidths $f_{\text{3dB}}$. In the narrow-band PN regime, the first-order BEM achieves the lowest BLER. This is because the PN is mainly dominated by its direct-current component and behaves approximately as a quasi-static phase rotation, for which the direct-current term alone is sufficient. As $f_{\text{3dB}}$ increases, however, the PN varies more rapidly over time, and the quasi-static assumption no longer holds. In this case, a higher-order BEM is required to capture the temporal evolution of the phase trajectory more accurately. Nevertheless, further increasing the BEM order, e.g., $Q \ge 3$, results in performance degradation rather than improvement. This phenomenon can be attributed to overfitting, where excessive basis functions introduce unnecessary degrees of freedom and tend to fit noise fluctuations rather than the underlying phase dynamics. Therefore, a proper choice of the BEM order is essential to balance modeling accuracy and decoding reliability.

\begin{figure}[htbp]
\centerline{\includegraphics[width=0.46\textwidth]{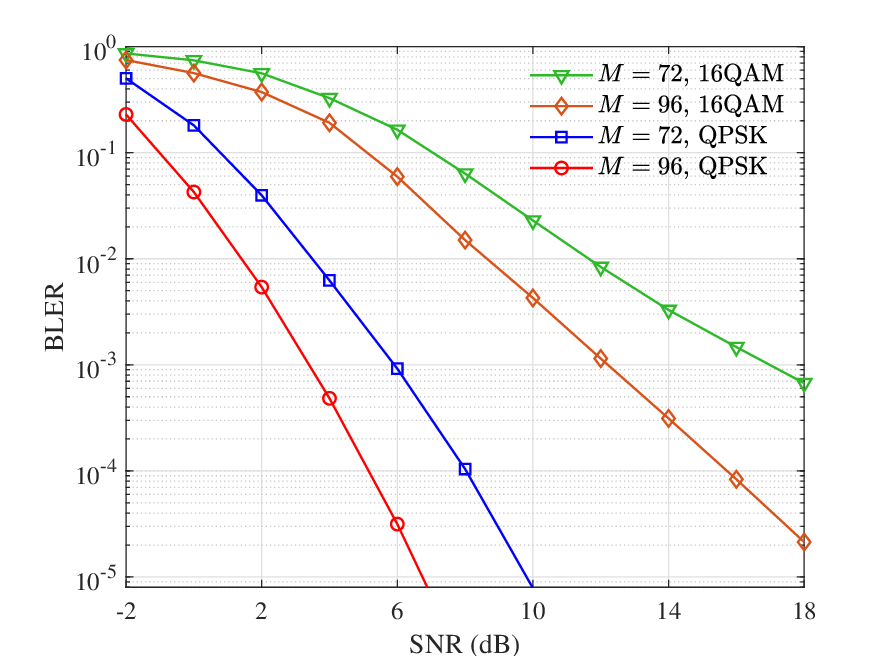}}
\caption{BLER of DM-SVC scheme under different
modulation orders and block lengths with $K_{\mathrm D}=2$, $K_{\mathrm P}=1$, $L_{\mathrm P}=2$.\vspace{-0pt}}
\end{figure}

Fig.~12 shows the BLER performance of the DM-SVC scheme under QPSK ($b=24$) and 16QAM ($b=28$ bits). It is observed that QPSK achieves lower BLER than 16QAM for both code lengths. The decision errors limit the effectiveness of the subsequent iterative PN refinement, which further degrades the decoding performance. In addition, the residual PN after BEM-based PN modeling and compensation leads to an error floor for 16-QAM in the high-SNR region. Compared with QPSK, the smaller minimum Euclidean distance of 16-QAM makes symbol detection more sensitive to residual phase distortion.
By contrast, QPSK is more robust to residual PN and thus enables more reliable iterative PN estimation and data decoding. Moreover, increasing the codeword length improves the BLER for both modulations, since a larger $M$ not only benefits sparse support recovery but also provides more observations for BEM coefficient estimation, thereby improving PN reconstruction accuracy. These results indicate that the modulation order and codeword length should be jointly designed in DM-SVC systems to balance transmission efficiency and decoding reliability.

Fig.~13 compares the BLER performance of proposed DM-SVC scheme with conventional SVC schemes in scenarios without PN. All schemes adopt the same codeword length $M = 96$ and QPSK modulation. In this PN-free scenario, the pilot symbols in DM-SVC are not required for PN estimation, and all non-zero elements and their indices are used for data transmission. For the DM-SVC scheme, two configurations are considered: case I with $K_{\mathrm P}=1$, $L_{\mathrm P}=2$, $K_{\mathrm D}=2$, $N=260$, and $b=31$, and case II with $K_{\mathrm P}=1$, $L_{\mathrm P}=3$, $K_{\mathrm D}=1$, $N=2100$, and $b=30$. For a fair comparison, the conventional schemes are configured with four non-zero elements and $b=30$, matching the total number of non-zero elements in the DM-SVC scheme. It can be observed that DM-SVC with case II achieves the best BLER performance among all compared schemes. This gain mainly benefits from its block-sparse structure. With the same total number of non-zero elements, the case II only needs to identify one non-zero block index and one non-zero element index, whereas the benchmark schemes need to detect four non-zero index, resulting in a larger support search space. Compared with case II, case I suffers from BLER degradation due to the enlarged index detection space, but it reduces the length of sparse vector, thereby reducing the codebook size and decoding complexity. This indicates that the proposed DM-SVC scheme offers additional flexibility in balancing transmission reliability and codebook size by adjusting the non-zero block length and the number of non-zero elements.

\begin{figure}[htbp]
\centerline{\includegraphics[width=0.46\textwidth]{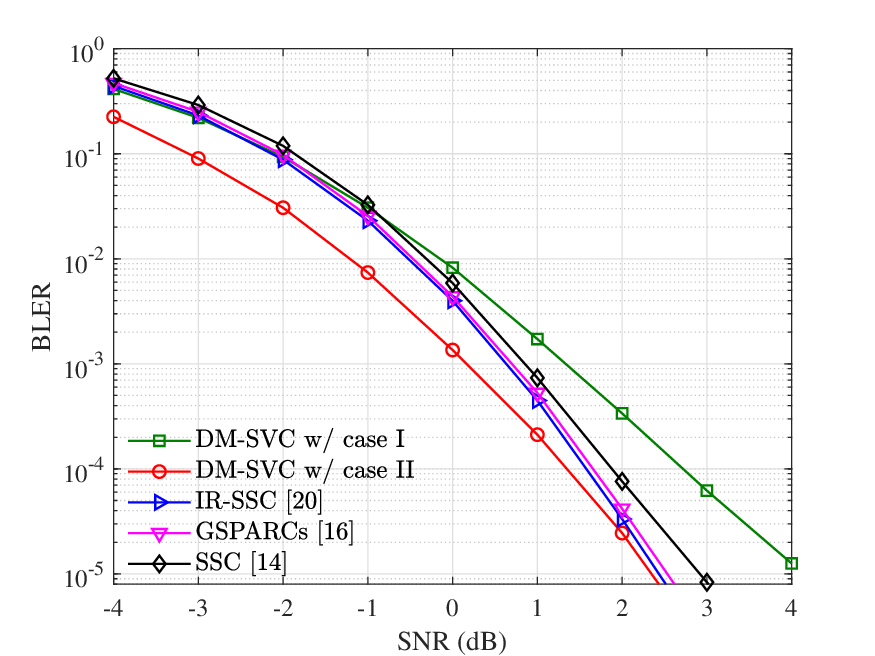}}\vspace{-0pt}
\caption{BLER performance comparison of different schemes without PN.\vspace{-0pt}}
\end{figure}
\begin{figure}[htbp]
\centerline{\includegraphics[width=0.46\textwidth]{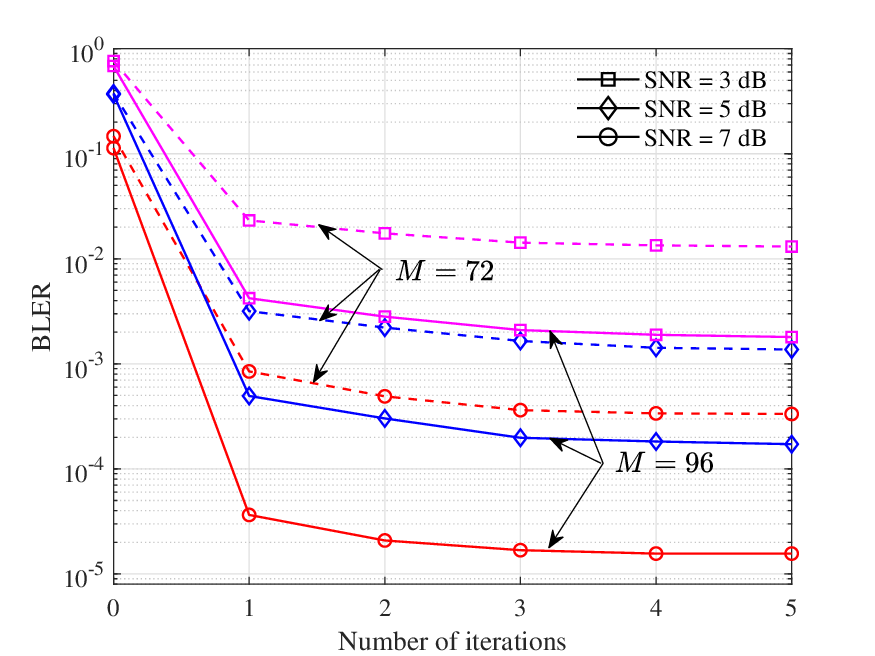}}\vspace{-0pt}
\caption{Convergence speed of proposed DM-SVC scheme with $K_{\mathrm D}=2$, $K_{\mathrm P}=1$, $L_{\mathrm P}=2$, $M=72$, and $b=24$.\vspace{-0pt}}
\end{figure}

\subsection{Convergence Performance}
Fig.~14 shows the convergence behavior of the proposed DM-SVC scheme for different SNR levels and code lengths. For all configurations, the BLER decreases monotonically with the number of iterations. A significant BLER reduction is observed within the first two iterations, and the performance stabilizes after approximately three iterations across all SNR conditions. This fast convergence is mainly due to the iterative interaction between PN estimation and data decoding. In particular, the decoded data symbols in previous iterations are used as pseudo-pilots to further refine the PN estimate. The improved PN estimate then enhances non-zero elements and symbol detection in the next iteration, forming a positive feedback process. This demonstrates that the proposed receiver achieves fast convergence with only a few iterations, making it well suited for low-latency short-packet transmission scenarios.

\begin{figure}[htbp]
\centerline{\includegraphics[width=0.46\textwidth]{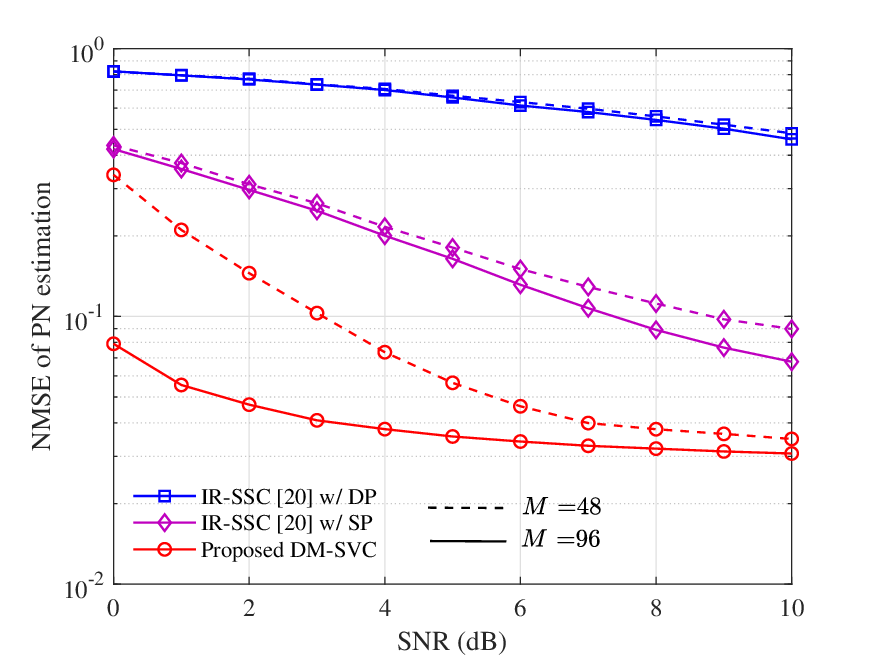}}
\caption{NMSE performance of PN estimation versus SNR for different schemes and code lengths with $K_{\mathrm D}=2$, $K_{\mathrm P}=1$, $L_{\mathrm P}=2$ and $b=24$.\vspace{-0pt}}
\end{figure}

\subsection{Performance of PN Estimation}

Fig. 15 compares the NMSE performance of PN estimation of proposed DM-SVC scheme with baseline schemes under different codeword lengths. It can be observed that DM-SVC achieves the lowest NMSE across the considered SNR range. The IR-SSC scheme with DP exhibits the worst PN estimation performance, as it neither exploits BEM-based PN modeling nor employs a data-assisted PN refinement strategy. Although the IR-SSC scheme with SP reuses data-bearing resources for PN estimation, its large correlation of codebook degrades decoding reliability, which in turn limits the accuracy of PN estimation. In contrast, the proposed DM-SVC scheme separates pilot and data components through distinct sparsity patterns and iteratively refines PN estimation with the aid of decoded data symbols, enabling more accurate PN estimation. Moreover, the PN estimation accuracy of DM-SVC exhibits sensitivity to the codeword length in the low-SNR regime. This is because PN estimation in the DM-SVC scheme relies on correct identification of non-zero pilot blocks. A longer codeword improves pilot block detection reliability, thereby reducing the PN estimation error. As the SNR increases, the NMSE gap among different codeword lengths gradually diminishes. This indicates that the performance gain from increasing the codeword length becomes marginal when AWGN is no longer the dominant impairment.


\vspace{-0pt}
\section{Conclusions}

This paper investigated the problem of PN impairment in SVC-based short-packet transmission systems. The proposed DM-SVC scheme integrates pilots and data into a unified sparse vector through a dual-mapping structure, thereby eliminating the need for dedicated pilot resources. In addition, the pilot block indices provide an extra dimension for information transmission, which improves the SE compared with conventional SVC schemes. By adopting BEM-based PN modeling, the PN estimation problem is reduced to the estimation of a small number of coefficients. Based on this model, the proposed iterative joint decoder converges within a few iterations. Simulation results showed that the proposed DM-SVC scheme achieves BLER performance close to the perfect-PN baseline while requiring lower codebook storage overhead than existing schemes. Future work will focus on reducing the decoding complexity and extending the proposed scheme to massive MIMO systems.

\vspace{-0pt}
\bibliographystyle{IEEEbib}
\bibliography{refs}

\end{document}